\documentclass[twocolumn,journal]{IEEEtran}
\usepackage[T1]{fontenc}
\usepackage[latin9]{inputenc}
\usepackage{float}
\usepackage{amsmath}
\usepackage{amsthm}
\usepackage{amssymb}
\usepackage{stackrel}
\usepackage{graphicx}
\usepackage[unicode=true,
 bookmarks=true,bookmarksnumbered=true,bookmarksopen=true,bookmarksopenlevel=1,
 breaklinks=false,pdfborder={0 0 0},pdfborderstyle={},backref=false,colorlinks=false]
 {hyperref}
\hypersetup{pdftitle={Your Title},
 pdfauthor={Your Name},
 pdfpagelayout=OneColumn, pdfnewwindow=true, pdfstartview=XYZ, plainpages=false}

\makeatletter

\floatstyle{ruled}
\newfloat{algorithm}{tbp}{loa}
\providecommand{\algorithmname}{Algorithm}
\floatname{algorithm}{\protect\algorithmname}

\theoremstyle{plain}
\newtheorem{thm}{\protect\theoremname}
\theoremstyle{remark}
\newtheorem{rem}[thm]{\protect\remarkname}
\theoremstyle{plain}
\newtheorem{lem}[thm]{\protect\lemmaname}

\usepackage[caption=false,font=footnotesize]{subfig}

\@ifundefined{showcaptionsetup}{}{%
 \PassOptionsToPackage{caption=false}{subfig}}
\usepackage{subfig}
\makeatother

\providecommand{\lemmaname}{Lemma}
\providecommand{\remarkname}{Remark}
\providecommand{\theoremname}{Theorem}

\begin{document}
\title{PDMM: A novel Primal-Dual Majorization-Minimization algorithm for
Poisson Phase-Retrieval problem}
\author{Ghania Fatima, Zongyu Li, Aakash Arora, and Prabhu Babu\thanks{Ghania Fatima and Prabhu Babu are with CARE, IIT Delhi, New Delhi,
India (e-mail: Ghania.Fatima@care.iitd.ac.in, prabhubabu@care.iitd.ac.in).}\thanks{Zongyu Li is with Department of Electrical Engineering and Computer
Science, University of Michigan, Ann Arbor, MI 48109-2122 (e-mail:
zonyul@umich.edu).}\thanks{Aakash Arora is with SnT, University of Luxembourg, Luxembourg (e-mail:
aakash.arora@uni.lu).}\thanks{Zongyu Li is funded by NSF grant IIS 1838179, and NIH grants R01 EB022075
and R01 CA240706.}}
\maketitle
\begin{abstract}
In this paper, we introduce a novel iterative algorithm for the problem
of phase-retrieval where the measurements consist of only the magnitude
of linear function of the unknown signal, and the noise in the measurements
follow Poisson distribution. The proposed algorithm is based on the
principle of majorization-minimization (MM); however, the application
of MM here is very novel and distinct from the way MM has been usually
used to solve optimization problems in the literature. More precisely,
we reformulate the original minimization problem into a saddle point
problem by invoking Fenchel dual representation of the $\mathrm{log}\left(.\right)$
term in the Poisson likelihood function. We then propose tighter surrogate
functions over both primal and dual variables resulting in a double-loop
MM algorithm, which we have named as Primal-Dual Majorization-Minimization
(PDMM) algorithm. The iterative steps of the resulting algorithm are
simple to implement and involve only computing matrix vector products.
We also extend our algorithm to handle various $\ell_{1}$ regularized
Poisson phase-retrieval problems (which exploit sparsity). The proposed
algorithm is compared with previously proposed algorithms such as
wirtinger flow (WF), MM (conventional), and alternating direction
methods of multipliers (ADMM) for the Poisson data model. The simulation
results under different experimental settings show that PDMM is faster
than the competing methods, and its performance in recovering the
original signal is at par with the state-of-the-art algorithms.
\end{abstract}

\begin{IEEEkeywords}
Phase-retrieval, Poisson data model, Majorization-Minimization (MM),
Saddle-point problem, Fenchel dual representation.
\end{IEEEkeywords}

\section{INTRODUCTION AND LITERATURE}

Many physical measurement systems measure only the magnitude (or magnitude
square) of the signal and not its phase. For example, optical devices
(e.g., CCD cameras, photosensitive films) cannot directly measure
the phase of the light wave because of the high rate of electromagnetic
field oscillations ($\sim10^{15}$ Hz) and measure only the photon
flux, which is the magnitude square of the electromagnetic field \cite{key-1}.
At a large enough distance from the imaging plane, the electromagnetic
field structure is given by the product of the Fourier transform of
the image and a known phase factor. It is known that in the Fourier
representation of signals, the magnitude and phase play different
roles \cite{key-2}, and in many cases, the important features of
the signal are preserved only in the phase. Since important phase
information is lost, the recorded signal does not resemble the original
signal, and recovering the phase from the Fourier magnitude is a challenging
task. Therefore, it is important to come up with sturdy algorithms
that successfully retrieve the original signal with only the magnitude
information at hand. The problem of recovering the original signal
from the magnitude (or magnitude square) of its linear measurements
is called phase-retrieval problem. Various fields in which the problem
of phase-retrieval arises are optical imaging \cite{key-3}, astronomical
imaging \cite{key-4}, speech and audio processing \cite{key-5,key-6,key-7,key-8},
crystallography \cite{key-9,key-10,key-11}, computational biology
\cite{key-12}, and electron microscopy \cite{key-13,key-14,key-15}.

Mathematically, phase retrieval is to obtain a $K$-dimensional complex
valued signal $\mathbf{x}$ from $N$ measurements denoted by $\mathbf{y}\in\mathbb{R_{+}^{\mathit{N}}}$,
which are nothing but magnitude squares of some linear function of
$\mathbf{x}$ and are modelled as:
\begin{equation}
\small y_{i}=\left|\boldsymbol{\mathrm{a}}_{i}^{\mathrm{H}}\mathbf{x}\right|^{2}+b_{i},\,i=1,...,N.\label{eq:1}
\end{equation}
The measurement vectors $\left\{ \mathbf{a}_{i}^{\mathrm{H}}\right\} _{i=1}^{N}\in\mathbb{C^{\mathit{K}}}$
that corresponds to the rows of the measurement matrix $\boldsymbol{\mathrm{A}}\in\mathbb{C^{\mathit{N\times K}}}$
(assumed to have full column rank), and the mean background signal
$b_{i}\in\mathbb{\mathbb{R}}_{+}$ for the $i^{\mathrm{th}}$ measurement
are usually known beforehand. In the cases where $y_{i}'\mathrm{s}$
correspond to magnitude square of the Fourier transform of $\mathbf{x}$,
$\mathbf{a}_{i}^{\mathrm{H}}$ would be the rows of Discrete Fourier
Transform (DFT) matrix. Often the measurements are corrupted with
noise and thus (\ref{eq:1}) becomes
\begin{equation}
\small y_{i}=\left|\boldsymbol{\mathrm{a}}_{i}^{\mathrm{H}}\mathbf{x}\right|^{2}+b_{i}+n_{i},\,i=1,...,N,
\end{equation}
 where $n_{i}$ denotes the measurement noise.

Since important phase information is lost, the number of measurements
($N$) should be in general larger than the dimension of the signal
($K$) for the stable recovery of the signal. The authors of \cite{key-16}
proved that $N$ should at least be of the order of $K\,\mathrm{log}\,K$
in order to successfully recover the original signal. Furthermore,
the authors of \cite{key-17} proved that $N=4\,K-4$ is necessary
and sufficient to uniquely recover the original signal.

Most of the previous works done on phase-retrieval assumes the data
model to be Gaussian where the entries of $\mathbf{y}$ are statistically
independent and follow Gaussian distribution, i.e.,
\begin{equation}
\small y_{i}\sim\mathcal{N}\left(\left|\boldsymbol{\mathrm{a}}_{i}^{\mathrm{H}}\mathbf{x}\right|^{2}+b_{i},\sigma^{2}\right),
\end{equation}
where $\sigma^{2}$ denotes the variance of the noise in the measurements.
The maximum likelihood (ML) estimate of $\mathbf{x}$ in this case
would be obtained by solving the following non-convex problem:
\begin{equation}
\small\underset{\mathbf{x}}{\mathrm{min}}\;\stackrel[i=1]{\mathrm{\mathit{N}}}{\sum}\left|y_{i}-b_{i}-\left|\boldsymbol{\mathrm{a}}_{i}^{\mathrm{H}}\mathbf{x}\right|^{2}\right|^{2}.\label{eq:4}
\end{equation}
A number of algorithms have been previously proposed to solve for
a minimizer of (\ref{eq:4}). One of the recent approaches involves
the reformulation of the phase-retrieval problem by a technique called
``matrix-lifting'' \cite{key-16,key-18,key-19}, wherein a rank-1
matrix $\mathbf{X}:=\mathbf{x}\mathbf{x}^{\mathrm{H}}$ is introduced
and the problem is transformed into a higher dimensional space making
the Fourier magnitude square measurements linear in $\mathbf{X}$.
Although the objective of the transformed problem is convex, the resulting
problem is non-convex due to the presence of rank-1 constraint, which
is then relaxed and the resultant problem is solved using semi-definite
programming (SDP). The SDP based algorithm yields robust solutions
but is computationally demanding, reducing their applications to only
low dimension problems. In the classical WF \cite{key-20} algorithm
for phase-retrieval, the step size is updated in an ad-hoc manner
to descend the cost function in (\ref{eq:4}). The step size is initially
kept small and then increased with each iteration, thereby requiring
a hyper parameter for its control. PRIME \cite{key-21} uses the MM
technique wherein a sequence of some surrogate problems are solved
instead of the original non-convex problem yielding a simple iterative
algorithm.

The authors of \cite{key-22,key-23,key-24} proposed algorithms for
phase-retrieval problem based on the magnitude model instead of the
intensity (or squared magnitude) model in (\ref{eq:1}). In \cite{key-22},
the authors have proposed an algorithm known as Gerchberg-Saxton (GS)
algorithm, wherein a new auxiliary variable is introduced to reformulate
the problem and the equivalent problem is then solved by alternating
minimization. In \cite{key-23}, a phase-retrieval algorithm using
ADMM was proposed wherein auxiliary magnitude and phase variables
are introduced to eliminate the absolute value operator. More recently,
the authors of \cite{key-24} proposed an iterative soft-thresholding
with exact line search algorithm (STELA) for sparse phase-retrieval
which is based on the successive convex approximation (SCA) framework.
The authors of \cite{key-25} combined the ideas of SDP relaxation
and GS algorithm to come up with a novel method that lifts phase-vector
instead of the original signal $\mathbf{x}$ and deduced convex relaxation
for the non-convex problem which was then solved using block coordinate
descent method.

Although the assumption of Gaussian noise in the measurements is common,
in low photon count applications such as coherent diffractive imaging
\cite{key-26}, ptychography \cite{key-27,key-28} and holographic
phase retrieval \cite{key-29}, the light source is weak and consequently,
the detection of SNR is limited by the quantized nature of light,
wherein the shot noise cannot be avoided \cite{key-30}. Therefore
in such cases, assuming Poisson distribution to model the noise is
more appropriate, i.e.,
\begin{equation}
\small y_{i}\sim\mathrm{Poisson}\left(\left|\boldsymbol{\mathrm{a}}_{i}^{\mathrm{H}}\mathbf{x}\right|^{2}+b_{i}\right).\label{eq:6}
\end{equation}
The ML estimate of $\mathbf{x}$ for the Poisson data model can be
obtained by solving the following non-convex problem:
\begin{equation}
\small\underset{\mathbf{x}}{\mathrm{min}}\;\stackrel[i=1]{\mathrm{\mathit{N}}}{\sum}\left[\left|\boldsymbol{\mathrm{a}}_{i}^{\mathrm{H}}\mathbf{x}\right|^{2}+b_{i}-y_{i}\,\mathrm{log}\,\left(\left|\boldsymbol{\mathrm{a}}_{i}^{\mathrm{H}}\mathbf{x}\right|^{2}+b_{i}\right)\right].\label{eq:7}
\end{equation}
In the works of \cite{key-31,key-32,key-33,key-34,key-35,key-36},
various authors have considered data models similar to (\ref{eq:6})
for phase-retrieval for the case of background signal $b_{i}=0$.
However, background signal is rarely zero in real world applications.
The authors of \cite{key-36} have proposed a variational model based
on Total Variation (TV) regularization for phase-retrieval problem
and an algorithm based on ADMM to solve a problem similar to (\ref{eq:7}).
The authors of \cite{key-37} have recently proposed three phase-retrieval
algorithms for solving (\ref{eq:7}); the three algorithms they proposed
are WF, MM and ADMM. In the WF algorithm for Poisson phase-retrieval,
instead of using a heuristic step size, the authors have proposed
a step size based on the observed Fisher information that can be computed
without any tuning parameter. In their MM based algorithm, quadratic
majorizer using improved curvature was used; however, this algorithm
does not work for cases where the background signal $b_{i}=0$. To
handle such cases, an algorithm based on ADMM was introduced which
is somewhat similar to the ADMM based phase-retrieval algorithm proposed
in \cite{key-36}.

In this paper, we propose a new algorithm for Poisson phase-retrieval
problem which is based on MM framework and it is different from the
MM based algorithm proposed in \cite{key-37}. The proposed algorithm
introduces an auxiliary variable into the original minimization problem
and converts it into a saddle-point minimax problem by invoking Fenchel
dual representation of the log term in the original cost function.
The modified problem is then solved using MM framework by proposing
surrogate functions over both the primal and dual variables to obtain
a saddle-point. The resulting double loop MM algorithm is named as
\textbf{P}rimal-\textbf{D}ual \textbf{M}ajorization-\textbf{M}inimization
(PDMM) algorithm and is referred by this name henceforth. The MM-based
algorithm proposed in \cite{key-37} does not work when the background
signal $b_{i}=0$, whereas PDMM effectively handles both $b_{i}>0$
and $b_{i}=0$ cases. Moreover, the surrogate function employed in
PDMM is numerically shown to be tighter to the objective function
in (\ref{eq:7}) than the quadratic surrogate function proposed in
\cite{key-37}. (Please refer to Fig. 1 and the related discussion
given in section \ref{subsec:3.2}).

The main contributions of this paper can be summarized as follows:
\begin{enumerate}
\item We have devised a novel primal-dual MM algorithm which is different
and efficient from the conventional MM method (which is restrictive
as it is applicable only for the case $b_{i}>0$) for Poisson phase-retrieval
problem.
\item We present the computationally efficient implementation of the proposed
algorithm requiring only simple matrix-vector product and Hadamard
product of vectors.
\item We discuss the extension of our PDMM algorithm for the regularized
Poisson phase-retrieval problem with the choice of regularizer $r(\mathbf{x})=\left\Vert \mathbf{T}\mathbf{x}\right\Vert _{1}$,
where unlike the restriction needed in the literature, the matrix
$\mathbf{T}$ may not be proximal friendly.
\item We discuss the proof of convergence of the proposed algorithm and
show that the algorithm always converges to a stationary point of
the Poisson likelihood function.
\item We present several numerical simulation results (for both one-dimensional
and two-dimensional signals) under different experimental settings
and make performance comparisons with previously proposed algorithms
(WF, conventional MM, and ADMM).
\end{enumerate}
The rest of the paper is organized as follows: Section II formulates
the minimization problem for Poisson phase-retrieval data model. Section
III gives a brief overview of the MM framework, introduces the proposed
algorithm, and discusses the computational complexity and the proof
of convergence. Section IV gives the numerical simulation details
and results under various experimental settings and comparison with
the results of state-of-the-art algorithms. Finally, section V concludes
the paper.

$\emph{Notation}:$ Bold upper case letters (e.g., $\mathbf{A}$ ,
$\mathbf{C}$) denotes matrices while bold lower case letters (e.g.,
\textbf{$\mathbf{x}$ },\textbf{ $\mathbf{y}$}) denotes column vectors.
Italics (e.g.,\textit{ $c$ , $d$}) denotes scalars. For a vector
$\mathbf{x}$,\textbf{ $\left|\mathbf{x}\right|$} denotes element-wise
magnitude, $x_{i}$ denotes the $i^{\mathrm{th}}$ element of the
vector,\textbf{ }$\left\Vert \mathbf{x}\right\Vert _{2}$ denotes
the Euclidean norm and $\mathbf{x}\circ\mathbf{y}$ denotes the Hadamard
product with vector $\mathbf{y}$. The superscript $\left(.\right)^{\ast}$denotes
the conjugate, and the superscript $\left(.\right)^{\mathrm{H}}$,
$\left(.\right)^{\mathrm{T}}$ and $\left(.\right)^{\dagger}$ denotes
the conjugate transpose, transpose and the pseudo-inverse of a matrix
respectively. For a complex number $x$, $\mathrm{Re}(x)$ and $\mathrm{Im}(x)$
denotes the real and imaginary parts respectively. $\mathbf{I}$ denotes
the Identity matrix. The symbol $\mathrm{diag}(\mathbf{x})$ denotes
the diagonal matrix formed by the elements of vector $\mathbf{x}$
as the principal diagonal and $\mathrm{diag}(\mathbf{X})$ denotes
the column vector formed by the diagonal elements of matrix $\mathbf{X}$.
The symbol $\mathrm{trace}(\mathbf{X})$ denotes the trace of the
matrix $\mathbf{X}$. The subscript $\mathbf{x}_{t}$ denotes the
vector \textbf{$\mathbf{x}$} at the $t^{\mathrm{th}}$ iteration,
and the vector $\mathbf{x}^{\mathrm{opt}}$ denotes the optimal value
of \textbf{$\mathbf{x}$}.

\section{PROBLEM FORMULATION}

In the case of Poisson data model, $\small y_{i}\sim\mathrm{Poisson}\left(\left|\boldsymbol{\mathrm{a}}_{i}^{\mathrm{H}}\mathbf{x}\right|^{2}+b_{i}\right),\:i=1,...,N$,
the ML estimation corresponds to solving the following minimization
problem:
\begin{equation}
\small\underset{\mathbf{x}}{\mathrm{min}}\;\left\{ f(\mathbf{x})\triangleq\stackrel[i=1]{\mathrm{\mathit{N}}}{\sum}\left[\left|\boldsymbol{\mathrm{a}}_{i}^{\mathrm{H}}\mathbf{x}\right|^{2}+b_{i}-y_{i}\,\mathrm{log}\,\left(\left|\boldsymbol{\mathrm{a}}_{i}^{\mathrm{H}}\mathbf{x}\right|^{2}+b_{i}\right)\right]\right\} .\label{eq:8}
\end{equation}
The function $f(\mathbf{x})$ is non-convex due to the presence of
quadratic term inside $-\mathrm{log\left(.\right)}$, which also makes
it a challenging optimization problem to solve. To devise the proposed
algorithm, we use the Fenchel representation of the log function \cite{key-38}
given as:
\begin{equation}
\small\mathrm{-log}(u)=\underset{z\geq0}{\mathrm{max}}\;\mathrm{log}(z)-zu+1
\end{equation}
Using the above mentioned representation, we get the following saddle-point
formulation which is equivalent to the original optimization problem:
\begin{equation}
\small\underset{\mathbf{x}}{\mathrm{min}}\:\underset{\mathbf{z}\geq0}{\mathrm{max}}\;\stackrel[i=1]{\mathrm{\mathit{N}}}{\sum}\left[\left|\boldsymbol{\mathrm{a}}_{i}^{\mathrm{H}}\mathbf{x}\right|^{2}+b_{_{i}}+y_{i}\,\mathrm{log}\,(z_{i})-z_{i}\,\left(\left|\boldsymbol{\mathrm{a}}_{i}^{\mathrm{H}}\mathbf{x}\right|^{2}+b_{i}\right)\right],\label{eq:10}
\end{equation}
where $\boldsymbol{\mathbf{z}}$ denotes the vector containing the
elements $\{z_{i}\}$. The equivalence of (\ref{eq:8}) and (\ref{eq:10})
can be proved as follows. Consider the inner maximization problem
over $z_{i}$:
\begin{equation}
\small\underset{\mathbf{z}\geq0}{\mathrm{max}}\;\stackrel[i=1]{\mathrm{\mathit{N}}}{\sum}\left[y_{i}\,\mathrm{log}\,(z_{i})-z_{i}\,\left(\left|\boldsymbol{\mathrm{a}}_{i}^{\mathrm{H}}\mathbf{x}\right|^{2}+b_{i}\right)\right].
\end{equation}
The above problem is separable in $z_{i}$. Therefore, a generic problem
in $z$ (independent of index ``$i$'') can be written as
\begin{equation}
\small\underset{z\geq0}{\mathrm{max}}\;y\,\mathrm{log}\,(z)-z\,\left(\left|\boldsymbol{\mathrm{a}}^{\mathrm{H}}\mathbf{x}\right|^{2}+b\right).
\end{equation}
Writing the KKT conditions for the above problem and solving for $z$,
we get
\begin{equation}
\small z^{\mathrm{opt}}=\frac{y}{\left|\boldsymbol{\mathrm{a}}^{\mathrm{H}}\mathbf{x}\right|^{2}+b}.
\end{equation}
Putting the optimal value $z_{i}^{\mathrm{opt}}$ back in (\ref{eq:10})
we get the following minimization problem
\begin{equation}
\small\begin{aligned}\underset{\mathbf{x}}{\mathrm{min}}\;\stackrel[i=1]{\mathrm{\mathit{N}}}{\sum}\;\left[\left|\boldsymbol{\mathrm{a}}_{i}^{\mathrm{H}}\mathbf{x}\right|^{2}+b_{_{i}}+y_{i}\,\mathrm{log}\,\left(\frac{y_{i}}{\left|\boldsymbol{\mathrm{a}}_{i}^{\mathrm{H}}\mathbf{x}\right|^{2}+b_{i}}\right)\right.\\
\left.-\frac{y_{i}}{\left|\boldsymbol{\mathrm{a}}_{i}^{\mathrm{H}}\mathbf{x}\right|^{2}+b_{i}}\,\left(\left|\boldsymbol{\mathrm{a}}_{i}^{\mathrm{H}}\mathbf{x}\right|^{2}+b_{i}\right)\right].
\end{aligned}
\end{equation}
Simplifying the above problem yields:
\begin{equation}
\small\underset{\mathbf{x}}{\mathrm{min}}\;\stackrel[i=1]{\mathrm{\mathit{N}}}{\sum}\left[\left|\boldsymbol{\mathrm{a}}_{i}^{\mathrm{H}}\mathbf{x}\right|^{2}+b_{_{i}}-y_{i}\,\mathrm{log}\,\left(\left|\boldsymbol{\mathrm{a}}_{i}^{\mathrm{H}}\mathbf{x}\right|^{2}+b_{i}\right)\right],
\end{equation}
which is same as problem (\ref{eq:8}). Thus the proof of the equivalence
between (\ref{eq:8}) and (\ref{eq:10}) is established.

By introducing an auxiliary variable $z_{i}$ (which can also be viewed
as a dual variable) in the original problem, the minimization problem
is converted into a minimax problem. A solution ($\mathbf{x}^{\mathrm{opt}},\mathbf{z}^{\mathrm{opt}}$)
of problem (\ref{eq:10}) would be a saddle-point for the minimax
problem and hence the reformulation in (\ref{eq:10}) can also be
termed as a saddle-point problem. Thus the crux of the proposed algorithm
is to utilize Fenchel representation to remove quadratic term inside
the $\mathrm{log}\left(.\right)$ and solve the equivalent saddle-point
problem (\ref{eq:10}) instead of the original problem (\ref{eq:8})
to arrive at an optimal solution for $\boldsymbol{\mathrm{x}}$.

The reader may doubt how the reformulation in (\ref{eq:10}), which
looks complicated (as it is a minimax problem) than the original problem
in (\ref{eq:8}), would be helpful, but in the next section we will
show clearly that it is indeed easy to work with (\ref{eq:10}) and
an MM algorithm can be devised. Moreover, the surrogate functions
devised in this case result in a much tighter approximation to the
original objective function in (\ref{eq:8}) than the surrogate function
employed by the MM algorithm developed in \cite{key-37}.

\section{PDMM: THE PROPOSED PHASE-RETRIEVAL ALGORITHM}

This section first gives a brief overview of the MM framework which
forms the backbone of the proposed algorithm. The proposed algorithms
for the unregularized and regularized phase-retrieval problem are
then explained later in detail. The section ends with a discussion
on computational complexity and the proof of convergence of the proposed
algorithm.

\subsection{\label{subsec:3.1}MM Framework \cite{key-45}}

For cases where the original minimization problem has a complicated
form, as in (\ref{eq:8}) and (\ref{eq:10}) (in case of (\ref{eq:10}),
the objective of the minimization problem also involves maximization
operation), MM exploits the problem structure and devises a problem-driven
algorithm. It works in two steps. The first step is the majorization
step, where a surrogate function which globally upperbounds the objective
function, with their difference minimized at the current point is
constructed. The second step is the minimization step, where the surrogate
function obtained in the first step is minimized.

Consider the following minimization problem:
\begin{equation}
\small\begin{aligned}\underset{\mathbf{x}}{\mathrm{min}\;}f(\mathbf{x})\\
\mathrm{subject}\;\mathrm{to} & \;\mathbf{x}\in\mathit{\mathcal{X}},
\end{aligned}
\label{eq:11}
\end{equation}
where $\mathcal{X}$ is a non-empty closed set in $\mathbb{R}^{n}$/$\mathbb{C}^{n}$,
and $f(\mathbf{x})$ is a continuous function. It is assumed that
$f(\mathbf{x})$ goes to infinity when $\left\Vert \mathbf{x}\right\Vert _{2}\rightarrow+\infty$.
In MM, a point $\mathbf{x}_{0}\in\mathcal{X}$ is initialized and
a series of feasible points $\boldsymbol{\mathrm{x}}_{t}$ is generated.
In the first step, a surrogate function $g(\mathbf{x}|\mathbf{x}_{t})$
at $\mathbf{x}_{t}$ is constructed satisfying the following properties:
\begin{equation}
\small g(\mathbf{x}|\mathbf{x}_{t})\geq f(\mathbf{x}),\forall\mathbf{x}\in\mathcal{X}\label{eq:12}
\end{equation}
and
\begin{equation}
\small g(\mathbf{x}_{t}|\mathbf{x}_{t})=f(\mathbf{x}_{t}).\label{eq:18}
\end{equation}
In the second step which is the minimization step, $\mathbf{x}$ is
updated as
\begin{equation}
\small\mathbf{x}_{t+1}=\mathrm{arg}\:\underset{\mathbf{x}\in\mathcal{X}}{\mathrm{min}}\;g(\mathbf{x}|\mathbf{x}_{t}).\label{eq:13}
\end{equation}
From (\ref{eq:12}), (\ref{eq:18}) and (\ref{eq:13}), the following
inequality is deduced:
\begin{equation}
\small f(\mathbf{x}_{t+1})\leq g(\mathbf{x}_{t+1}|\mathbf{x}_{t})\leq g(\mathbf{x}_{t}|\mathbf{x}_{t})=f(\mathbf{x}_{t})\label{eq:14}
\end{equation}
which shows that the sequence $(f(\mathbf{x}_{t}))$ is non-increasing.
Thus, the objective function decreases monotonically using MM. The
success of MM lies in the appropriate formulation of surrogate function.
If the surrogate function is smooth and convex, and separable in variables,
its minimization becomes efficient and scalable, leading to algorithms
that can be easily implemented. However, while formulating a surrogate
function, there is a trade-off between faster convergence and low
computational cost/memory requirements per iteration. For faster convergence,
the surrogate function should follow the shape of the objective function
as close as possible. Whereas, to keep the computational complexity
low, the surrogate function should be simple to minimize. Balancing
this trade-off is the key to the successful implementation of MM.

Here we would like to mention that the general discussion for the
MM framework presented in this subsection is for a minimization problem
(as in (\ref{eq:11})), however our problem of interest (\ref{eq:10})
is a minimax problem. The next subsection explains how the above mentioned
steps of the MM can be adapted for a minimax problem.

\subsection{\label{subsec:3.2}Primal-Dual Majorization-Minimization (PDMM) Algorithm}

Let us start with the cost function (only over $\mathbf{x}$) in (\ref{eq:10}),
\begin{equation}
\small\begin{aligned}f(\mathbf{x})\triangleq\underset{\mathbf{z}\geq0}{\mathrm{max}}\;h(\mathbf{x},\mathbf{z})\triangleq\underset{\mathbf{z}\geq0}{\mathrm{max}}\;\left(\mathbf{x}^{\mathrm{H}}\mathbf{A}^{\mathrm{H}}\mathbf{A}\mathbf{x}+\stackrel[i=1]{\mathrm{\mathit{N}}}{\sum}y_{i}\,\mathrm{log\,}(z_{i})\right.\\
\left.-\stackrel[i=1]{\mathrm{\mathit{N}}}{\sum}z_{i}\mathbf{x}^{\mathrm{H}}\boldsymbol{\mathrm{a}}_{i}\boldsymbol{\mathrm{a}}_{i}^{\mathrm{H}}\mathbf{x}-\stackrel[i=1]{\mathrm{\mathit{N}}}{\sum}z_{i}b_{i}\right).
\end{aligned}
\label{eq:15}
\end{equation}
Then the optimization problem given in (\ref{eq:10}) can be rewritten
as:
\begin{equation}
\small\underset{\mathbf{x}}{\mathrm{min}}\:f(\mathbf{x})\label{eq:16}
\end{equation}

\noindent The third term of $h(\mathbf{x},\mathbf{z})$ in (\ref{eq:15}),
i.e., $-\stackrel[i=1]{\mathrm{\mathit{N}}}{\sum}z_{i}\boldsymbol{\mathrm{x}}^{\mathrm{H}}\boldsymbol{\mathrm{a}}_{i}\boldsymbol{\mathrm{a}}_{i}^{\mathrm{H}}\mathbf{x}$
is a concave function in $\mathbf{x}$ (for any $z_{i}$). Thus, we
can use first-order Taylor's expansion to upperbound the concave term,
which will be discussed shortly.

\noindent Let the third term of $h(\mathbf{x},\mathbf{z})$ be denoted
as:
\begin{equation}
\small h_{3}(\mathbf{x},\mathbf{z})\triangleq-\stackrel[i=1]{\mathrm{\mathit{N}}}{\sum}z_{i}\mathbf{x}^{\mathrm{H}}\boldsymbol{\mathrm{a}}_{i}\boldsymbol{\mathrm{a}}_{i}^{\mathrm{H}}\mathbf{x},
\end{equation}
which we can also express as follows: 
\begin{equation}
\small h_{3}(\mathbf{x},\mathbf{z})=\stackrel[i=1]{\mathrm{\mathit{N}}}{\sum}z_{i}\tilde{h}_{3i}(\mathbf{x}),
\end{equation}
where $\tilde{h}_{3i}(\mathbf{x})=-\boldsymbol{\mathrm{x}}^{\mathrm{H}}\boldsymbol{\mathrm{a}}_{i}\boldsymbol{\mathrm{a}}_{i}^{\mathrm{H}}\mathbf{x}$,
which is clearly a concave function in $\mathbf{x}$. Using the first-order
Taylor's expansion for $\tilde{h}_{3i}(\mathbf{x})$ at any $\mathbf{x}=\mathbf{x}_{\mathrm{\mathit{t}}},$
\begin{equation}
\small\tilde{h}_{3i}(\mathbf{x})\leq\tilde{h}_{3i}(\mathbf{x}_{t})+\mathrm{Re}\left(\nabla\tilde{h}_{3i}(\mathbf{x}_{t})^{\mathrm{H}}(\mathbf{x}-\mathbf{x}_{t})\right),\label{eq:19}
\end{equation}
we get the following MM inequality:
\begin{flalign}
\small-\mathbf{x}^{\mathrm{H}}\boldsymbol{\mathrm{a}}_{i}\boldsymbol{\mathrm{a}}_{i}^{\mathrm{H}}\mathbf{x}\leq & \:\mathbf{x}_{t}^{\mathrm{H}}\boldsymbol{\mathrm{a}}_{i}\boldsymbol{\mathrm{a}}_{i}^{\mathrm{H}}\mathbf{x}_{t}-2\:\mathrm{Re}\left(\mathbf{x}_{t}^{\mathrm{H}}\boldsymbol{\mathrm{a}}_{i}\boldsymbol{\mathrm{a}}_{i}^{\mathrm{H}}\mathbf{x}\right),\label{eq:20}
\end{flalign}
where the equality is achieved when $\mathbf{x}=\mathbf{x}_{t}.$
Thus, using (\ref{eq:20}) $h_{3}(\mathbf{x},\mathbf{z})$ can be
upperbounded as
\begin{equation}
\small\begin{aligned}-\stackrel[i=1]{\mathrm{\mathit{N}}}{\sum}z_{i}\mathbf{x}^{\mathrm{H}}\boldsymbol{\mathrm{a}}_{i}\boldsymbol{\mathrm{a}}_{i}^{\mathrm{H}}\mathbf{x}\leq\;\stackrel[i=1]{\mathrm{\mathit{N}}}{\sum}z_{i}\mathbf{x}_{t}^{\mathrm{H}}\boldsymbol{\mathrm{a}}_{i}\boldsymbol{\mathrm{a}}_{i}^{\mathrm{H}}\mathbf{x}_{t}-2\stackrel[i=1]{\mathrm{\mathit{N}}}{\sum}\mathrm{Re}\left(z_{i}\mathbf{x}_{t}^{\mathrm{H}}\boldsymbol{\mathrm{a}}_{i}\boldsymbol{\mathrm{a}}_{i}^{\mathrm{H}}\mathbf{x}\right).\end{aligned}
\label{eq:21}
\end{equation}
Using (\ref{eq:21}), the objective function in (\ref{eq:15}) can
be upperbounded as
\begin{equation}
\small\begin{aligned}f(\mathbf{x})=\underset{\mathbf{z}\geq0}{\mathrm{max}}\;h(\mathbf{x},\mathbf{z})\leq\underset{\mathbf{z}\geq0}{\mathrm{max}}\;\left(\mathbf{x}^{\mathrm{H}}\mathbf{A}^{\mathrm{H}}\mathbf{A}\mathbf{x}-2\stackrel[i=1]{\mathrm{\mathit{N}}}{\sum}\mathrm{Re}\left(z_{i}\mathbf{x}_{t}^{\mathrm{H}}\boldsymbol{\mathrm{a}}_{i}\boldsymbol{\mathrm{a}}_{i}^{\mathrm{H}}\mathbf{x}\right)\right.\\
\left.+\stackrel[i=1]{\mathrm{\mathit{N}}}{\sum}z_{i}\mathbf{x}_{t}^{\mathrm{H}}\boldsymbol{\mathrm{a}}_{i}\boldsymbol{\mathrm{a}}_{i}^{\mathrm{H}}\mathbf{x}_{t}+\stackrel[i=1]{\mathrm{\mathit{N}}}{\sum}y_{i}\,\mathrm{log\,}(z_{i})-\stackrel[i=1]{\mathrm{\mathit{N}}}{\sum}z_{i}b_{i}\right)\triangleq g_{f}(\mathbf{x}|\mathbf{x}_{t}).
\end{aligned}
\end{equation}
Thus, we arrive at the following surrogate optimization problem:
\begin{equation}
\small\begin{aligned}\underset{\mathbf{x}}{\mathrm{min}}\:\left\{ g_{f}(\mathbf{x}|\mathbf{x}_{t})=\underset{\mathbf{z}\geq0}{\mathrm{max}}\;\left(\mathbf{x}^{\mathrm{H}}\mathbf{A}^{\mathrm{H}}\mathbf{A}\mathbf{x}+\stackrel[i=1]{\mathit{\mathrm{\mathit{N}}}}{\sum}y_{i}\,\mathrm{log}\,(z_{i})-\stackrel[i=1]{\mathrm{\mathit{N}}}{\sum}z_{i}b_{i}\right.\right.\\
\left.\left.+\stackrel[i=1]{\mathit{\mathrm{\mathit{N}}}}{\sum}z_{i}\left|d_{i}\right|^{2}-2\stackrel[i=1]{\mathrm{\mathit{N}}}{\sum}\mathrm{Re}\left(z_{i}d_{i}^{\ast}\boldsymbol{\mathrm{a}}_{i}^{\mathrm{H}}\mathbf{x}\right)\right)\right\} ,
\end{aligned}
\label{eq:23}
\end{equation}
where $d_{i}\triangleq\mathbf{a}_{i}^{\mathrm{H}}\mathbf{x}_{t}$.
It can be noted that the above problem has an inner maximization problem
in $\mathbf{z}$ and an outer maximization problem in $\mathbf{x}.$
If we solve the inner maximization problem and substitute back the
optimal maximizer $\mathbf{z}^{\mathrm{opt}}$, we get a minimization
problem only in the primal variable $\mathbf{x}$, in which the objective
(in $\mathbf{x}$) would be a MM surrogate for the objective in (\ref{eq:8}).
In the following remark we will explain more on this aspect.
\begin{rem}
The inner maximization problem of (\ref{eq:23}) can be written as:
\begin{equation}
\small\begin{aligned}\underset{\mathbf{z}\geq0}{\mathrm{max}}\:\stackrel[i=1]{\mathit{\mathrm{\mathit{N}}}}{\sum}\left[y_{i}\,\mathrm{log}\,(z_{i})-z_{i}b_{i}+z_{i}\left|d_{i}\right|^{2}-2\mathrm{Re}\left(z_{i}d_{i}^{\ast}\boldsymbol{\mathrm{a}}_{i}^{\mathrm{H}}\mathbf{x}\right)\right].\end{aligned}
\label{eq:24}
\end{equation}
It is obvious from the objective in (\ref{eq:24}) that for any $i$,
when $b_{i}-\left|d_{i}\right|^{2}+2\mathrm{Re}\left(d_{i}^{\ast}\mathbf{a}_{i}^{\mathrm{H}}\mathbf{x}\right)\leq0$,
the optimization problem in (\ref{eq:24}) would be unbounded above,
and when $b_{i}-\left|d_{i}\right|^{2}+2\mathrm{Re}\left(d_{i}^{\ast}\mathbf{a}_{i}^{\mathrm{H}}\mathbf{x}\right)>0$,
the optimal value of $z_{i}$ would be
\begin{equation}
\small z_{i}^{\mathrm{opt}}=\frac{y_{i}}{b_{i}-\left|d_{i}\right|^{2}+2\mathrm{Re}\left(d_{i}^{\mathrm{\ast}}\boldsymbol{\mathrm{a}}_{i}^{\mathrm{H}}\mathbf{x}\right)}.
\end{equation}
Substituting back $z_{i}^{\mathrm{opt}}$ in (\ref{eq:23}) (with
the condition that $b_{i}-\left|d_{i}\right|^{2}+2\mathrm{Re}\left(d_{i}^{\ast}\mathbf{a}_{i}^{\mathrm{H}}\mathbf{x}\right)>0\;\forall i$),
we get the following surrogate problem of PDMM in primal variable
$\mathbf{x}:$
\begin{equation}
\small\begin{aligned}\underset{\mathbf{x}}{\mathrm{min}}\;\left(\mathbf{x}^{\mathrm{H}}\mathbf{A}^{\mathrm{H}}\mathbf{A}\mathbf{x}-\stackrel[i=1]{\mathrm{\mathit{N}}}{\sum}\left(\frac{b_{i}y_{i}}{b_{i}-\left|d_{i}\right|^{2}+2\mathrm{Re}\left(d_{i}^{\mathrm{\ast}}\boldsymbol{\mathrm{a}}_{i}^{\mathrm{H}}\mathbf{x}\right)}\right)\right.\\
+\stackrel[i=1]{\mathit{\mathrm{\mathit{N}}}}{\sum}y_{i}\,\mathrm{log}\,\left(\frac{y_{i}}{b_{i}-\left|d_{i}\right|^{2}+2\mathrm{Re}\left(d_{i}^{\mathrm{\ast}}\boldsymbol{\mathrm{a}}_{i}^{\mathrm{H}}\mathbf{x}\right)}\right)\\
-2\stackrel[i=1]{\mathrm{\mathit{N}}}{\sum}\mathrm{Re}\left(\frac{y_{i}d_{i}^{\mathrm{\ast}}\boldsymbol{\mathrm{a}}_{i}^{\mathrm{H}}\mathbf{x}}{b_{i}-\left|d_{i}\right|^{2}+2\mathrm{Re}\left(d_{i}^{\ast}\boldsymbol{\mathrm{a}}_{i}^{\mathrm{H}}\mathbf{x}\right)}\right)\\
\left.+\stackrel[i=1]{\mathit{\mathrm{\mathit{N}}}}{\sum}\left(\frac{y_{i}\left|d_{i}\right|^{2}}{b_{i}-\left|d_{i}\right|^{2}+2\mathrm{Re}\left(d_{i}^{\mathrm{\ast}}\boldsymbol{\mathrm{a}}_{i}^{\mathrm{H}}\mathbf{x}\right)}\right)\right)\\
\mathrm{subject}\mathrm{\;to}\;b_{i}-\left|d_{i}\right|^{2}+2\mathrm{Re}\left(d_{i}^{\mathrm{\ast}}\boldsymbol{\mathrm{a}}_{i}^{\mathrm{H}}\mathbf{x}\right)>0\;\forall i
\end{aligned}
\label{eq:28}
\end{equation}
The objective in (\ref{eq:28}) is a surrogate function for the original
Poisson likelihood function in (\ref{eq:8}). For instance, consider
a one-dimensional example with two data samples ($N=2$), the elements
of $\mathbf{A}\in\mathbb{R}^{1\times2}$ randomly generated, $b_{i}=0.1,$
and $\mathbf{y}$ generated via the model in (\ref{eq:6}) with the
choice of true $x$ ($x_{\mathrm{true}}$) fixed to $8.$ Fig. 1 shows
the plots of the objective in (\ref{eq:8}) around the neighbourhood
of $x_{\mathrm{true}}$, the objective of (\ref{eq:28}) satisfying
the constraint, and also the ``MM'' surrogate function developed
in \cite{key-37} with the choice of $x_{t}=4$. From Fig. 1 it can
be seen that both the function of (\ref{eq:28}) and the quadratic
majorizer of \cite{key-37} exactly match at $x=4$ (as required in
MM), however the function in (\ref{eq:28}) closely follows the shape
of the original Poisson likelihood function and hence is a tighter
approximation to it than the one proposed in \cite{key-37}, and it
is more likely to yield a faster converging algorithm if we solve
(\ref{eq:28}), then iterate and so on. However, (\ref{eq:28}) does
not have a closed form solution and requires a convex solver like
CVX to solve it, so we proceed further by reformulating the minimax
problem (\ref{eq:23}) into a maximin problem and look for other possibilities
to solve the problem. Nonetheless, Fig.1 clearly shows that the surrogate
constructed in PDMM is a tighter approximation to the original objective
than the surrogate proposed in \cite{key-37}.
\end{rem}
\begin{figure}[tbh]
\begin{centering}
\includegraphics[width=6.8cm,height=3.6cm]{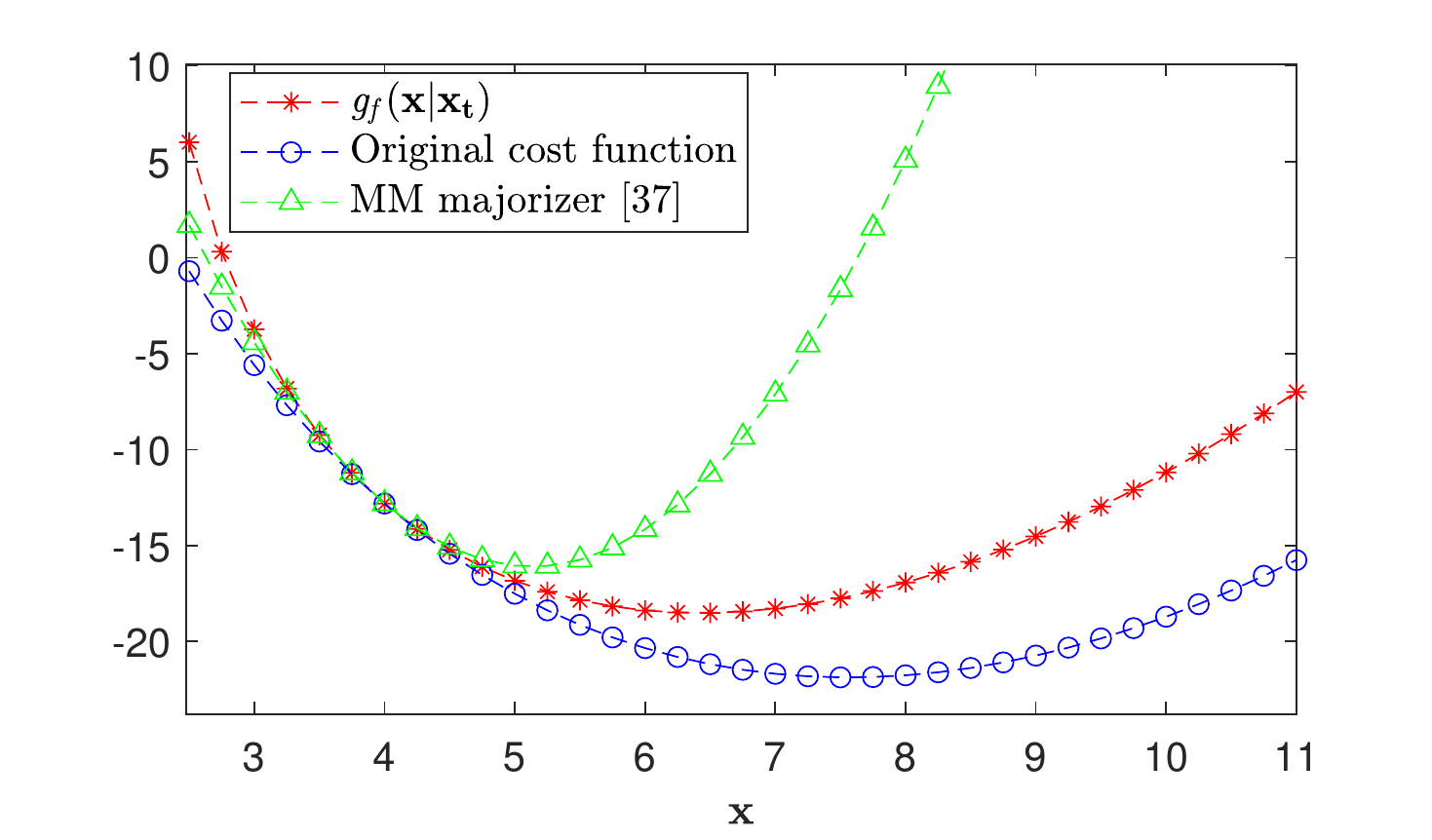}
\par\end{centering}
\caption{A representative plot of majorizers for PDMM (satisfying the constraint
in (\ref{eq:28})) and MM \cite{key-37} with $x_{t}=4$ for the non-convex
Poisson log-likelihood function.}
\end{figure}

\noindent To proceed further, we first rewrite the optimization problem
in (\ref{eq:23}) as:
\begin{equation}
\begin{aligned}\underset{\mathbf{x}}{\mathrm{min}}\:\underset{\mathbf{z}\geq0}{\mathrm{max}}\;\left(\mathbf{x}^{\mathrm{H}}\mathbf{A}^{\mathrm{H}}\mathbf{A}\mathbf{x}+\stackrel[i=1]{\mathit{\mathrm{\mathit{N}}}}{\sum}y_{i}\,\mathrm{log}\,(z_{i})-\stackrel[i=1]{\mathrm{\mathit{N}}}{\sum}z_{i}b_{i}\right.\\
\left.+\stackrel[i=1]{\mathit{\mathrm{\mathit{N}}}}{\sum}z_{i}\left|d_{i}\right|^{2}-2\stackrel[i=1]{\mathrm{\mathit{N}}}{\sum}\mathrm{Re}\left(z_{i}d_{i}^{\ast}\boldsymbol{\mathrm{a}}_{i}^{\mathrm{H}}\mathbf{x}\right)\right)
\end{aligned}
\label{eq:33-1}
\end{equation}
The minimax problem in (\ref{eq:33-1}) can be reformulated as a maximin
problem using minimax theorem \cite{key-39} (which is briefly stated
in the following lemma):
\begin{lem}
\label{lem:2}Let $\mathcal{X}\subset\mathbb{R}^{n}$and $\mathcal{Y}\subset\mathbb{R}^{m}$
be compact convex sets . If $f:\mathcal{X}\times\mathcal{Y}\rightarrow\mathbb{R}$
is a continuous function that is concave-convex, i.e., $f(.,y):\mathcal{X}\rightarrow\mathbb{R}$
is concave for fixed y, and $f(x,.):\mathcal{Y}\rightarrow\mathbb{R}$
is convex for fixed x, then we have 
\[
\small\underset{x\in\mathcal{X}}{\mathrm{max}}\:\underset{y\in\mathcal{Y}}{\mathrm{min}}\:f(x,y)=\underset{y\in\mathcal{Y}}{\mathrm{min}}\:\underset{x\in\mathcal{X}}{\mathrm{max}}\:f(x,y)
\]
\end{lem}
\noindent Although the variable $\mathbf{x}\in\mathbb{C}^{K}$ is
complex valued, if we rewrite (\ref{eq:33-1}) using a new variable
$\tilde{\mathbf{x}}\triangleq\left[\mathrm{Re}\left(\mathbf{x}\right);\mathrm{Im}\left(\mathbf{x}\right)\right]\in\mathbb{R}^{2K}$,
the objective function in (\ref{eq:33-1}) would still be a continuous
function which is convex in $\tilde{\mathbf{x}}$ for fixed $\mathbf{z}$
and concave in $\mathbf{z}$ for fixed $\tilde{\mathbf{x}}$. Therefore
Lemma \ref{lem:2} can be applied to the minimax problem in (\ref{eq:33-1}).
Thus, swapping min and max in (\ref{eq:23}), we get
\begin{equation}
\small\begin{aligned}\underset{\mathbf{z}\geq0}{\mathrm{max}}\:\underset{\mathbf{x}}{\mathrm{min}}\;\left(\mathbf{x}^{\mathrm{H}}\mathbf{A}^{\mathrm{H}}\mathbf{A}\mathbf{x}-2\,\mathrm{Re}\left(\stackrel[i=1]{\mathrm{\mathit{N}}}{\sum}\left(z_{i}d_{i}^{\mathrm{\ast}}\mathbf{a}_{i}^{\mathrm{H}}\boldsymbol{\mathrm{x}}\right)\right)-\stackrel[i=1]{\mathrm{\mathit{N}}}{\sum}z_{i}b_{i}\right.\\
\left.+\stackrel[i=1]{\mathrm{\mathit{N}}}{\sum}y_{i}\,\mathrm{log}\,(z_{i})+\stackrel[i=1]{\mathrm{\mathit{N}}}{\sum}z_{i}\left|d_{i}\right|^{2}\right).
\end{aligned}
\label{eq:29}
\end{equation}
The maximin problem (\ref{eq:29}) has an inner minimization problem
in \textbf{$\boldsymbol{\mathrm{x}}$} and an outer maximization problem
in $\mathbf{z}$. We first solve the inner minimization problem which
is given by:
\begin{equation}
\small\underset{\boldsymbol{\mathrm{x}}}{\mathrm{min}}\;\mathbf{x}^{\mathrm{H}}\mathbf{A}^{\mathrm{H}}\mathbf{A}\mathbf{x}-2\,\mathrm{Re}\left(\stackrel[i=1]{\mathrm{\mathit{N}}}{\sum}\left(z_{i}d_{i}^{\mathrm{\ast}}\mathbf{a}_{i}^{\mathrm{H}}\boldsymbol{\mathrm{x}}\right)\right).
\end{equation}
The minimizer over \textbf{$\boldsymbol{\mathrm{x}}$} is given as:
\begin{equation}
\small\mathbf{x}_{t+1}=\left(\mathbf{A}^{\mathrm{H}}\mathbf{A}\right)^{-1}\mathbf{A}^{\mathrm{H}}\mathbf{Dz}\label{eq:31}
\end{equation}
where $\mathbf{D}=\mathrm{diag}(\mathbf{d})$ i.e. a diagonal matrix
formed by setting elements of $\mathbf{d}$ as its principal diagonal.
It is to be noted that $\left(\mathbf{A}^{\mathrm{H}}\mathbf{A}\right)^{-1}\mathbf{A}^{\mathrm{H}}$
is the pseudo-inverse of matrix $\mathbf{A}$ and hence (\ref{eq:31})
can be compactly written as:
\begin{equation}
\small\mathbf{x}_{t+1}=\mathbf{A}^{\dagger}(\mathbf{d}\circ\mathbf{z}).\label{eq:32}
\end{equation}
The pseudo-inverse $\mathbf{A}^{\dagger}$ is not iteration dependent
and can be calculated and stored (for suitably small problems) once
outside the iteration loops which will help us to implement the algorithm
in an efficient manner. By substituting back $\mathbf{x}_{t+1}$ in
(\ref{eq:29}) we get
\begin{equation}
\small\begin{aligned}\underset{\mathbf{z}\geq0}{\mathrm{max}}\;\left(-\mathbf{z}^{\mathrm{H}}\boldsymbol{\mathrm{D}}^{\mathrm{H}}\mathbf{A}\left(\mathbf{A}^{\mathrm{H}}\mathbf{A}\right)^{-1}\mathbf{A}^{\mathrm{H}}\mathbf{Dz}+\stackrel[i=1]{\mathit{\mathit{\mathrm{\mathit{N}}}}}{\sum}y_{i}\,\mathrm{log}\,(z_{i})\right.\\
\left.-\stackrel[i=1]{\mathrm{\mathit{N}}}{\sum}z_{i}b_{i}+\stackrel[i=1]{\mathrm{\mathit{N}}}{\sum}z_{i}\left|d_{i}\right|^{2}\right).
\end{aligned}
\label{eq:33}
\end{equation}

\noindent It is to be noted here that although the vector $\mathbf{z}$
is real-valued, we use $\mathbf{z}^{\mathrm{H}}$ to denote its transpose
for consistency in the notation. The maximization problem (\ref{eq:33})
is first converted into the following minimization problem for convenience:
\begin{equation}
\small\begin{aligned}\underset{\mathbf{z}\geq0}{\mathrm{min}}\:\mathbf{z}^{\mathrm{H}}\boldsymbol{\mathrm{D}}^{\mathrm{H}}\mathbf{A}\left(\mathbf{A}^{\mathrm{H}}\mathbf{A}\right)^{-1}\mathbf{A}^{\mathrm{H}}\mathbf{Dz}-\stackrel[i=1]{\mathrm{\mathit{N}}}{\sum}y_{i}\,\mathrm{log}\,(z_{i})\\
+\stackrel[i=1]{\mathrm{\mathit{N}}}{\sum}z_{i}b_{i}-\stackrel[i=1]{\mathrm{\mathit{N}}}{\sum}z_{i}\left|d_{i}\right|^{2}
\end{aligned}
\label{eq:34}
\end{equation}

\noindent The above mentioned problem is a convex minimization problem
in $\{z_{i}\}'s$ and can be solved directly using interior point
solver like CVX \cite{key-40}. However, it would be computationally
inefficient to use CVX (especially for large dimensional problem setting).
Therefore, in the following we explore the use of MM (over variables
$z_{i}$) again to find the global minimizer of the problem in (\ref{eq:34}).
We construct a surrogate function for the objective in (\ref{eq:34})
at some given $\boldsymbol{\mathbf{z}}_{k}$ and minimize the surrogate
iteratively to arrive at the optimal minimizer of (\ref{eq:34}).

\noindent Let $\boldsymbol{\mathrm{P}}\triangleq\mathbf{A}\left(\mathbf{A}^{\mathrm{H}}\mathbf{A}\right)^{-1}\mathbf{A}^{\mathrm{H}}$
and $\mathbf{I}$ denote an identity matrix of dimension $N\times N$.
The minimization problem (\ref{eq:34}) can be rewritten as:
\begin{equation}
\small\begin{aligned}\underset{\mathbf{z}\geq0}{\mathrm{min}}\;\left(\boldsymbol{\mathrm{z}}^{\mathrm{H}}\boldsymbol{\mathrm{D}}^{\mathrm{H}}(\boldsymbol{\mathrm{P}-}\boldsymbol{\mathrm{I}})\boldsymbol{\mathrm{D}}\boldsymbol{\mathrm{z}}+\boldsymbol{\mathrm{z}}^{\mathrm{H}}\mathbf{D}^{\mathrm{H}}\mathbf{D}\boldsymbol{\mathrm{z}}+\stackrel[i=1]{\mathrm{\mathit{N}}}{\sum}z_{i}b_{i}\right.\\
\left.-\stackrel[i=1]{\mathit{\mathit{\mathrm{\mathit{N}}}}}{\sum}y_{i}\,\mathrm{log}\,(z_{i})-\stackrel[i=1]{\mathrm{\mathit{N}}}{\sum}z_{i}\left|d_{i}\right|^{2}\right).
\end{aligned}
\label{eq:35}
\end{equation}

\begin{rem}
It is to be noted that $\mathrm{\mathbf{P}}$ is a projection matrix
onto the column space of matrix $\mathbf{A}$. The matrix $\mathbf{I-P}$
is then the projection matrix onto its orthogonal complement or null
space of $\mathbf{A}^{\mathrm{H}}$. Therefore, matrices $\mathbf{P}$
and $\mathbf{I-P}$ are Hermitian, positive semi-definite, and have
a maximum eigenvalue equal to 1. The matrix $\mathbf{P-I}$ is therefore
a negative semi-definite matrix which makes the first term in problem
(\ref{eq:35}) concave.
\end{rem}
The concave term in (\ref{eq:35}) is linearized (via first-order
Taylor series expansion, similar to (\ref{eq:19}), at $\mathbf{z}=\mathbf{z}_{k}$)
and an upperbound is obtained, using which we arrive at the following
surrogate problem:
\begin{equation}
\small\begin{aligned}\underset{\mathbf{z}\geq0}{\mathrm{min}}\;\left(2\mathrm{Re}\left(\boldsymbol{\mathrm{z}}^{\mathrm{H}}\boldsymbol{\mathrm{D}}^{\mathrm{H}}(\boldsymbol{\mathrm{P}-}\boldsymbol{\mathrm{I}})\mathrm{\mathbf{D}\mathbf{z}_{\mathit{k}}}\right)+\boldsymbol{\mathrm{z}}^{\mathrm{H}}\boldsymbol{\mathbf{D}}^{\mathrm{H}}\mathbf{D}\boldsymbol{\mathrm{z}}+\stackrel[i=1]{\mathrm{\mathit{N}}}{\sum}z_{i}b_{i}\right.\\
\left.-\stackrel[i=1]{\mathit{\mathit{\mathrm{\mathit{N}}}}}{\sum}y_{i}\,\mathrm{log}\,(z_{i})-\stackrel[i=1]{\mathrm{\mathit{N}}}{\sum}z_{i}\left|d_{i}\right|^{2}\right).
\end{aligned}
\end{equation}
With $\mathbf{c}\triangleq2\mathrm{Re}\left(\mathbf{D}^{\mathrm{H}}(\boldsymbol{\mathrm{P}-}\mathbf{I})\mathbf{D}\boldsymbol{\mathrm{z}}_{k}\right)$
and $h_{i}\triangleq\left|d_{i}\right|^{2}$ , the above mentioned
problem becomes:
\begin{equation}
\small\begin{aligned}\underset{\mathbf{z}\geq0}{\mathrm{min}}\;\stackrel[i=1]{\mathrm{\mathit{N}}}{\sum}\left[c_{i}z_{i}+h_{i}z_{i}^{2}+b_{i}z_{i}-y_{i}\,\mathrm{log}\,(z_{i})-h_{i}z_{i}\right]\end{aligned}
.
\end{equation}
which is separable in $z_{i}$. Thus, a generic problem (without the
index $i$) can be written as
\begin{equation}
\small\underset{z\geq0}{\mathrm{min}}\;cz+hz^{2}+bz-y\,\mathrm{log}\,(z)-hz.\label{eq:38}
\end{equation}
The Karush-Kuhn-Tucker (KKT) condition for (\ref{eq:38}) will be
\begin{equation}
\small c+2hz+b-\frac{y}{z}-h=0,
\end{equation}
which can also be written as:
\begin{equation}
\small cz+2hz^{2}+bz-y-hz=0.\label{eq:40}
\end{equation}
Solving (\ref{eq:40}), we arrive at the optimal solution over $z$
which will serve as the next iterate:
\begin{equation}
\small z_{k+1}=\begin{cases}
\frac{-(b+c-h)+\sqrt{(b+c-h)^{2}+8hy}}{4h} & \mathrm{if}\;h\neq0\\
\frac{y}{b+c} & \mathrm{if}\;h=0
\end{cases}\label{eq:41}
\end{equation}
The pseudo code for the PDMM algorithm obtained is given in the Algorithm
Table \ref{alg:1}.
\begin{algorithm}[tbh]
\caption{\label{alg:1}Pseudo code of PDMM}

Input: $\boldsymbol{\mathrm{A}}$, $\boldsymbol{\mathrm{y}}$, $\boldsymbol{\mathrm{b}}$,
$\eta_{1}$ and $\eta_{2}$
\begin{enumerate}
\item Initialize $\mathbf{x}_{0}$ and $\mathbf{z}_{0}$
\item Compute $\mathbf{P}$ and $\mathbf{A}^{\dagger}$
\item Iterate: Given $\mathbf{x}_{t}$ , do the $(t+1)^{\mathrm{th}}$ step
\begin{enumerate}
\item Compute $\mathbf{d}$
\item Iterate: Given $\mathbf{z}_{k}$ , do the $(k+1)^{\mathrm{th}}$ step:
\begin{itemize}
\item Apply (\ref{eq:41}) to obtain $\mathbf{z}_{k+1}$
\item If $\left\Vert \mathbf{z}_{k+1}-\mathbf{z}_{k}\right\Vert _{2}/\left\Vert \mathbf{z}_{k}\right\Vert _{2}<\eta_{1},$
stop and return $\mathbf{z}_{k+1}.$
\end{itemize}
\item Apply (\ref{eq:32}) to obtain $\mathbf{x}_{t+1}$
\item If $\left\Vert \mathbf{x}_{t+1}-\mathbf{x}_{t}\right\Vert _{2}/\left\Vert \mathbf{x}_{t}\right\Vert _{2}<\eta_{2}$,
stop and return $\mathbf{x}_{t+1}.$
\end{enumerate}
\item $\mathbf{x}^{\mathrm{opt}}$ is the value of $\mathbf{x}$ returned
at convergence of outer loop.
\end{enumerate}
\end{algorithm}

\begin{rem}
\label{rem:4}The MM iterations over the variable $\mathbf{z}$ need
not necessarily be run till convergence. The inner loop iterations
can be stopped adaptively by computing $\mathbf{x}_{t+1}$ in (\ref{eq:32})
using the recent value of $\mathbf{z}$ and checking if $f(\mathbf{x}_{t+1})\leq f(\mathbf{x}_{t})$.
\end{rem}

\subsection{PDMM for regularized Poisson phase-retrieval problem}

The previous subsection describes the algorithm for minimizing the
unregularized problem, which is an ill-posed problem, especially when
$N<K$, so for the stable recovery of the original signal, one requires
some prior knowledge on the signal. One of the ways to deal with this
problem is to impose sparsity (which the underlying signal would also
possess) by incorporating an additional term (regularizer or penalty)
in the cost function \cite{key-41,key-42,key-43}. In this subsection,
we adapt the derivation for the regularized version of the Poisson
log-likelihood problem using the $\ell_{1}$ norm regularizer which
is well known to produce a sparse solution.

Let $r(\mathbf{x})$ be the regularizer, then the regularized problem
would be:
\begin{equation}
\small\phi(\mathbf{x})=f(\mathbf{x})+\lambda\,r(\mathbf{x})
\end{equation}
where $\lambda$ is the regularization parameter and $r(\mathbf{x})=\left\Vert \mathbf{T}\mathbf{x}\right\Vert _{1}$.
The different choices for $\mathbf{T}$ matrix can be Identity matrix,
Orthogonal Discrete Wavelet Transform (ODWT) matrix or the finite-difference
matrix used in Total variation (TV) regularization.

Using $r(\mathbf{x})=\left\Vert \mathbf{T}\mathbf{x}\right\Vert _{1}$,
the saddle point representation of the regularized problem can be
written as:
\begin{equation}
\small\begin{aligned}\underset{\mathbf{x}}{\mathrm{min}}\;\underset{\mathbf{z}\geq0}{\mathrm{max}}\;\left(\mathrm{\mathbf{x}^{H}\mathbf{A}^{H}\mathbf{A}\mathbf{x}}-\stackrel[i=1]{N}{\sum}z_{i}\mathbf{x}^{\mathrm{H}}\mathbf{a}_{i}\mathbf{a}_{i}^{\mathrm{H}}\mathbf{x}+\stackrel[i=1]{N}{\sum}y_{i}\mathrm{log}(z_{i})\right.\\
\left.-\stackrel[i=1]{N}{\sum}z_{i}b_{i}+\lambda\left\Vert \mathbf{T}\mathbf{x}\right\Vert _{1}\right).
\end{aligned}
\label{eq:43}
\end{equation}
Now, introducing one more auxiliary variable $w_{l}$'s to tackle
the regularization term, we get the following equivalent problem:
\begin{equation}
\small\begin{aligned}\underset{\mathbf{x}}{\mathrm{min}}\;\underset{\underset{|w_{l}|\leq1}{\mathbf{z}\geq0,}}{\mathrm{max}}\;\left(\mathrm{\mathbf{x}^{H}\mathbf{A}^{H}\mathbf{A}\mathbf{x}}-\stackrel[i=1]{N}{\sum}z_{i}\mathbf{x}^{\mathrm{H}}\mathbf{a}_{i}\mathbf{a}_{i}^{\mathrm{H}}\mathbf{x}+\stackrel[i=1]{N}{\sum}y_{i}\mathrm{log}(z_{i})\right.\\
\left.-\stackrel[i=1]{N}{\sum}z_{i}b_{i}+\lambda\mathrm{Re}\left(\mathbf{w}^{\mathrm{H}}\mathbf{Tx}\right)\right),
\end{aligned}
\label{eq:44}
\end{equation}
where $\mathbf{w}$ denotes the column vector containing $\{w_{l}\}_{l=1}^{K-1}$.
It is easy to verify that if we compute the maximizer over $w_{l}'s$
and substitute back the maximizer in the objective function in (\ref{eq:44}),
we will obtain the problem in (\ref{eq:43}). Following the similar
procedure as in the unregularized case, linearizing the concave term
$-\stackrel[i=1]{\mathrm{\mathit{N}}}{\sum}z_{i}\boldsymbol{\mathrm{x}}^{\mathrm{H}}\boldsymbol{\mathrm{a}}_{i}\boldsymbol{\mathrm{a}}_{i}^{\mathrm{H}}\mathbf{x}$
using first order Taylor expansion as given in (\ref{eq:19}), we
get the following surrogate problem.
\begin{equation}
\small\begin{aligned}\underset{\mathbf{x}}{\mathrm{min}}\:\underset{\underset{|w_{l}|\leq1}{\mathbf{z}\geq0,}}{\mathrm{max}}\;\left(\mathrm{\mathbf{x}^{H}\mathbf{A}^{H}\mathbf{A}\mathbf{x}}-2\mathrm{Re}\left(\stackrel[i=1]{N}{\sum}z_{i}d_{i}^{\ast}\mathbf{a}_{i}^{\mathrm{H}}\mathbf{x}\right)+\stackrel[i=1]{N}{\sum}\left|d_{i}\right|^{2}z_{i}\right.\\
\left.+\stackrel[i=1]{N}{\sum}y_{i}\mathrm{log}(z_{i})-\stackrel[i=1]{N}{\sum}z_{i}b_{i}+\lambda\mathrm{Re}\left(\mathbf{w}^{\mathrm{H}}\mathbf{Tx}\right)\right).
\end{aligned}
\end{equation}
Since the above mentioned problem is convex in $\mathbf{x}$ for fixed
$\mathbf{z}$ and $\mathbf{w}$, concave in $\mathbf{z}$ for fixed
$\mathbf{x}$ and $\mathbf{w}$, and linear in $\mathbf{w}$ for fixed
$\mathbf{x}$ and $\mathbf{z}$, the minmax theorem stated in Lemma
\ref{lem:2} can be applied on it in two steps. Swapping the $\underset{\mathbf{x}}{\mathrm{min}}$
and $\underset{\mathbf{z}\geq0}{\mathrm{max}}$ terms, we get the
intermediate problem:
\begin{equation}
\small\begin{aligned}\underset{\mathbf{z}\geq0}{\mathrm{max}}\:\underset{\mathbf{x}}{\mathrm{min}}\:\underset{w_{l}\leq1}{\mathrm{max}}\;\left(\mathrm{\mathbf{x}^{H}\mathbf{A}^{H}\mathbf{A}\mathbf{x}}-2\mathrm{Re}\left(\stackrel[i=1]{N}{\sum}z_{i}d_{i}^{\mathrm{\ast}}\mathbf{a}_{i}^{\mathrm{H}}\mathbf{x}\right)+\stackrel[i=1]{N}{\sum}\left|d_{i}\right|^{2}z_{i}\right.\\
\left.+\stackrel[i=1]{N}{\sum}y_{i}\mathrm{log(}z_{i})-\stackrel[i=1]{N}{\sum}z_{i}b_{i}+\lambda\mathrm{Re}\left(\mathbf{w}^{\mathrm{H}}\mathbf{Tx}\right)\right).
\end{aligned}
\end{equation}
In the second step, we swap $\underset{\mathbf{x}}{\mathrm{min}}$
and $\underset{w_{l}\leq1}{\mathrm{max}}$ to get the following maximin
optimization problem:
\begin{equation}
\small\begin{aligned}\underset{\underset{|w_{l}|\leq1}{\mathbf{z}\geq0,}}{\mathrm{max}}\;\underset{\mathbf{x}}{\mathrm{min}}\;\left(\mathrm{\mathbf{x}^{H}\mathbf{A}^{H}\mathbf{A}\mathbf{x}}-2\mathrm{Re}\left(\stackrel[i=1]{N}{\sum}z_{i}d_{i}^{\mathrm{\ast}}\mathbf{a}_{i}^{\mathrm{H}}\mathbf{x}\right)+\stackrel[i=1]{N}{\sum}\left|d_{i}\right|^{2}z_{i}\right.\\
\left.+\stackrel[i=1]{N}{\sum}y_{i}\mathrm{log(}z_{i})-\stackrel[i=1]{N}{\sum}z_{i}b_{i}+\lambda\mathrm{Re}\left(\mathbf{w}^{\mathrm{H}}\mathbf{Tx}\right)\right).
\end{aligned}
\label{eq:47}
\end{equation}
Problem (\ref{eq:47}) has an inner maximization problem in $\mathbf{x}$
and an outer maximization problem in $\mathbf{z}$ and $\mathbf{w}$.
We first solve the inner minimization problem which is given by:
\begin{equation}
\small\underset{\mathbf{x}}{\mathrm{min}}\;\mathrm{\mathbf{x}^{H}\mathbf{A}^{H}\mathbf{A}\mathbf{x}-2\mathrm{Re}\left(\stackrel[i=1]{N}{\sum}z_{i}d_{i}^{\mathrm{\ast}}\mathbf{a}_{i}^{\mathrm{H}}\mathbf{x}\right)+\lambda\mathrm{Re}\left(\mathbf{w}^{\mathrm{H}}\mathbf{Tx}\right).}
\end{equation}
The optimal solution for $\mathbf{x}$ would be
\begin{equation}
\small\mathbf{x}_{t+1}=\left(\mathbf{A}^{\mathrm{H}}\mathbf{A}\right){}^{-1}\left(\mathbf{A}^{\mathrm{H}}\mathbf{D}\mathbf{z}\boldsymbol{-}\frac{\lambda}{2}\mathbf{T}^{\mathrm{H}}\mathbf{w}\right),
\end{equation}
which can also be written as
\begin{equation}
\small\mathbf{x}_{t+1}=\mathbf{A}^{\dagger}\left(\mathbf{d}\circ\mathbf{z}\right)-\frac{\lambda}{2}\left(\mathbf{A}^{\mathrm{H}}\mathbf{A}\right){}^{-1}\left(\mathbf{T}^{\mathrm{H}}\mathbf{w}\right).\label{eq:50}
\end{equation}
The matrices $\mathbf{A}^{\dagger}$ and $\left(\mathbf{A}\mathbf{A}^{\mathrm{H}}\right){}^{-1}\mathbf{T}^{\mathrm{H}}$
in (\ref{eq:50}) can be pre-computed and stored since they are not
iteration dependent. Substituting back $\mathbf{x}_{t+1}$ in (\ref{eq:47})
gives the following maximization problem:
\begin{equation}
\small\begin{aligned}\underset{\underset{|w_{l}|\leq1}{\mathbf{z}\geq0,}}{\mathrm{max}}\left(-\mathrm{\mathbf{z}^{H}\mathbf{D}^{H}\mathbf{A}\left(\mathbf{\mathbf{A}^{\mathrm{H}}A}\right){}^{-1}\mathbf{A}^{H}\mathbf{Dz}}+\stackrel[i=1]{N}{\sum}\left|d_{i}\right|^{2}z_{i}+\stackrel[i=1]{N}{\sum}y_{i}\mathrm{log}(z_{i})\right.\\
+\frac{\lambda}{2}\mathbf{z}^{\mathrm{H}}\mathbf{D}^{\mathrm{H}}\mathbf{A}\left(\mathbf{A}^{\mathrm{H}}\mathbf{A}\right){}^{-1}\mathbf{T}^{\mathrm{H}}\mathbf{w}+\frac{\lambda}{2}\mathbf{w}^{\mathrm{H}}\mathbf{T}\left(\mathbf{A}^{\mathrm{H}}\mathbf{A}\right){}^{-1}\mathbf{A}^{\mathrm{H}}\mathbf{Dz}\\
\left.-\frac{\lambda^{2}}{4}\mathbf{w}^{\mathrm{H}}\mathbf{T}\left(\mathbf{A}^{\mathrm{H}}\mathbf{A}\right){}^{-1}\mathbf{T}^{\mathrm{H}}\mathbf{w}-\stackrel[i=1]{N}{\sum}z_{i}b_{i}\right).
\end{aligned}
\label{eq:51}
\end{equation}
We first rewrite the above problem into a minimization problem for
convenience.
\begin{equation}
\small\begin{aligned}\underset{\underset{|w_{l}|\leq1}{\mathbf{z}\geq0,}}{\mathrm{min}}\left(\mathrm{\mathbf{z}^{H}\mathbf{D}^{H}\mathbf{A}\left(\mathbf{\mathbf{A}^{\mathrm{H}}A}\right){}^{-1}\mathbf{A}^{H}\mathbf{Dz}}-\stackrel[i=1]{N}{\sum}\left|d_{i}\right|^{2}z_{i}-\stackrel[i=1]{N}{\sum}y_{i}\mathrm{log}(z_{i})\right.\\
+\stackrel[i=1]{N}{\sum}z_{i}b_{i}-\lambda\mathrm{Re}\left(\mathbf{z}^{\mathrm{H}}\mathbf{D}^{\mathrm{H}}\mathbf{A}\left(\mathbf{A}^{\mathrm{H}}\mathbf{A}\right){}^{-1}\mathbf{T}^{\mathrm{H}}\mathbf{w}\right)\\
\left.+\frac{\lambda^{2}}{4}\mathbf{w}^{\mathrm{H}}\mathbf{T}\left(\mathbf{A}^{\mathrm{H}}\mathbf{A}\right){}^{-1}\mathbf{T}^{\mathrm{H}}\mathbf{w}\right).
\end{aligned}
\label{eq:52}
\end{equation}
The above problem can be reformulated as a jointly convex problem
in $\mathbf{w}$ and $\mathbf{z}$ and can be solved using an interior
point solver. However, similar to the previous sub-section, we proceed
further to solve for $\mathbf{w}$ and $\mathbf{z}$ iteratively using
MM. With $\mathbf{X}\triangleq\left(\mathbf{A}^{\mathrm{H}}\mathbf{A}\right){}^{-1}\mathbf{T}^{\mathrm{H}}$
and $e$ denoting the maximum eigenvalue of $\mathbf{X}$ (which can
be pre-computed and stored as it is not iteration dependent), problem
(\ref{eq:52}) can be rewritten as:
\begin{equation}
\small\begin{aligned}\underset{\underset{|w_{l}|\leq1}{\mathbf{z}\geq0,}}{\mathrm{max}}\left(\boldsymbol{\mathrm{z}}^{\mathrm{H}}\mathrm{\mathbf{D}^{H}}(\boldsymbol{\mathrm{P}-}\boldsymbol{\mathrm{I}})\boldsymbol{\mathrm{Dz}}+\boldsymbol{\mathrm{z}}^{\mathrm{H}}\mathrm{\mathbf{D}^{H}}\mathbf{D}\boldsymbol{\mathrm{z}}+\stackrel[i=1]{N}{\sum}\left[z_{i}b_{i}-\left|d_{i}\right|^{2}z_{i}\right]\right.\\
-\stackrel[i=1]{N}{\sum}y_{i}\mathrm{log}(z_{i})-\lambda\mathrm{Re}\left(\mathbf{z}^{\mathrm{H}}\mathbf{D}^{\mathrm{H}}\mathbf{A}\left(\mathbf{A}^{\mathrm{H}}\mathbf{A}\right){}^{-1}\mathbf{T}^{\mathrm{H}}\mathbf{w}\right)\\
\left.+\frac{\lambda^{2}}{4}\mathbf{w}^{\mathrm{H}}\left(\boldsymbol{\mathrm{X}-}e\boldsymbol{\mathrm{I}}\right)\mathbf{w}+\frac{\lambda^{2}}{4}e\left\Vert \mathbf{w}\right\Vert _{2}^{2}\right).
\end{aligned}
\label{eq:53}
\end{equation}
The terms $\small\boldsymbol{\mathrm{z}}^{\mathrm{H}}\mathrm{\mathbf{D}^{H}}(\boldsymbol{\mathrm{P}-}\boldsymbol{\mathrm{I}})\boldsymbol{\mathrm{Dz}}$
and $\small\frac{\lambda^{2}}{4}\mathbf{w}^{\mathrm{H}}(\boldsymbol{\mathrm{X}-}e\boldsymbol{\mathrm{I}})\mathbf{w}$
in (\ref{eq:53}) are concave functions in $\mathbf{z}$ and $\mathbf{w}$,
respectively and therefore are linearized to give the following surrogate
minimization problem (the surrogate objective would be tighter at
($\boldsymbol{\mathbf{z}}_{k}$, $\mathbf{w}_{k}$)):
\begin{equation}
\small\begin{aligned}\underset{\underset{|w_{l}|\leq1}{\mathbf{z}\geq0,}}{\mathrm{max}}\;\left(2\mathrm{Re}\left(\mathbf{z}_{k}^{\mathrm{H}}\mathrm{\mathbf{D}^{H}}(\boldsymbol{\mathrm{P}-}\boldsymbol{\mathrm{I}})\boldsymbol{\mathrm{Dz}}\right)+\boldsymbol{\mathrm{z}}^{\mathrm{H}}\mathrm{\mathbf{D}^{H}}\mathbf{D}\boldsymbol{\mathrm{z}}-\stackrel[i=1]{N}{\sum}z_{i}\left|d_{i}\right|^{2}\right.\\
-\stackrel[i=1]{N}{\sum}y_{i}\mathrm{log}(z_{i})-\lambda\mathrm{Re}\left(\mathbf{z}^{\mathrm{H}}\mathrm{\mathbf{D}^{H}}\mathbf{A}\left(\mathbf{A}^{\mathrm{H}}\mathbf{A}\right){}^{-1}\mathbf{T}^{\mathrm{H}}\mathbf{w}\right)\\
\left.+\frac{\lambda^{2}}{2}\mathrm{Re}\left(\mathbf{w}_{k}^{\mathrm{H}}(\boldsymbol{\mathrm{X}-}e\mathrm{\boldsymbol{I}})\mathbf{w}\right)+\frac{\lambda^{2}}{4}e\left\Vert \mathbf{w}\right\Vert _{2}^{2}+\stackrel[i=1]{N}{\sum}z_{i}b_{i}\right).
\end{aligned}
\label{eq:54}
\end{equation}
Problem (\ref{eq:54}) can be solved by alternatingly minimizing it
with respect to $\mathbf{z}$ and $\mathbf{w}$ \cite{key-44}. Keeping
$\mathbf{w}_{k}$ fixed, we first solve for solve for $\mathbf{z}_{k+1}$
as follows:
\begin{equation}
\small\begin{aligned}\underset{\mathbf{z}\geq0}{\mathrm{min}}\;\left(2\mathrm{Re}\left(\mathbf{z}_{k}^{\mathrm{H}}\mathbf{D}^{\mathrm{H}}(\boldsymbol{\mathrm{P}-}\boldsymbol{\mathrm{I}})\boldsymbol{\mathrm{Dz}}\right)+\stackrel[i=1]{N}{\sum}\left[z_{i}b_{i}-z_{i}\left|d_{i}\right|^{2}-y_{i}\mathrm{log}(z_{i})\right]\right.\\
\left.+\boldsymbol{\mathrm{z}}^{\mathrm{H}}\mathbf{D}^{\mathrm{H}}\mathbf{D}\boldsymbol{\mathrm{z}}-\lambda\mathrm{Re}\left(\mathbf{z}^{\mathrm{H}}\mathbf{D}^{\mathrm{H}}\mathbf{A}\left(\mathbf{A}^{\mathrm{H}}\mathbf{A}\right){}^{-1}\mathbf{T}^{\mathrm{H}}\mathbf{w}_{k}\right)\right).
\end{aligned}
\end{equation}
With $\small\mathbf{c}\triangleq2\mathrm{Re}\left(\mathbf{D}^{\mathrm{H}}(\boldsymbol{\mathrm{P}-}\mathbf{I})\mathbf{D}\boldsymbol{\mathrm{z}}_{k}\right)$
and $\mathbf{\small g}\triangleq\lambda\mathrm{Re}\left(\mathbf{w}_{k}^{\mathrm{H}}\mathbf{T}(\mathbf{A}^{\mathrm{H}}\mathbf{A})^{-1}\mathbf{\mathbf{A}^{\mathrm{H}}}\mathbf{D}\right)$,
the minimization problem over $\mathbf{z}$ becomes separable in $z_{i}$
as shown below:
\begin{equation}
\small\begin{split}\underset{z_{i}\geq0}{\mathrm{min}}\;\stackrel[i=1]{N}{\sum}\left[c_{i}z_{i}+h_{i}z_{i}^{2}-g_{i}z_{i}+z_{i}b_{i}-y_{i}\mathrm{log}(z_{i})-h_{i}z_{i}\right].\end{split}
\end{equation}
Thus a generic problem in one variable (without the index $i$) can
be written as 
\begin{equation}
\small\underset{z\geq0}{\mathrm{min}}\;cz+hz^{2}-gz+zb-y\mathrm{log}(z)-hz.
\end{equation}
Writing KKT condition and solving it we get the optimal solution as:
\begin{equation}
\small z_{k+1}=\begin{cases}
\frac{-(b+c-g-h)+\sqrt{(b+c-g-h)^{2}+8hy}}{4h}, & \mathrm{if}\;h\neq0\\
\frac{y}{b+c-g}, & \mathrm{if}\;h=0
\end{cases}\label{eq:58}
\end{equation}
Now for a fixed $\mathbf{z}_{k+1}$, we solve for $\mathbf{w}_{k+1}$
as follows:
\begin{equation}
\small\begin{split}\underset{|w_{l}|\leq1}{\mathrm{min}}\;\left(\frac{\lambda^{2}}{2}\mathrm{Re}\left(\mathbf{w}_{k}^{\mathrm{H}}(\boldsymbol{\mathrm{X}-}e\boldsymbol{\mathrm{I}})\mathbf{w}\right)+\frac{\lambda^{2}}{4}e\left\Vert \mathbf{w}\right\Vert _{2}^{2}\right.\\
-\lambda\mathrm{Re}\left(\mathrm{\mathbf{z}}_{k+1}^{\mathrm{H}}\mathbf{D}^{\mathrm{H}}\mathbf{A}\left(\mathbf{A}^{\mathrm{H}}\mathbf{A}\right){}^{-1}\mathbf{T}^{\mathrm{H}}\mathbf{w}\right)\biggr).
\end{split}
\label{eq:59}
\end{equation}
By scaling out the factor $\frac{\lambda^{2}}{4}e$, we get:
\begin{equation}
\small\begin{aligned}\underset{|w_{l}|\leq1}{\mathrm{min}}\;\left\Vert \mathbf{w}\right\Vert _{2}^{2}-2\mathrm{Re}\left(\frac{2}{\lambda e}\mathrm{\mathbf{z}}_{k+1}^{\mathrm{H}}\mathbf{D}^{\mathrm{H}}\mathbf{A}\left(\mathbf{A}^{\mathrm{H}}\mathbf{A}\right){}^{-1}\mathbf{T}^{\mathrm{H}}\mathbf{w}\right.\\
-\frac{1}{e}\mathbf{w}_{k}^{\mathrm{H}}\left(\boldsymbol{\mathrm{X}-}e\boldsymbol{\mathrm{I}}\right)\mathbf{w}\biggr).
\end{aligned}
\end{equation}
With $\small\mathbf{u}\triangleq\frac{2}{\lambda e}\mathbf{T}\left(\mathbf{A}^{\mathrm{H}}\mathbf{A}\right){}^{-1}\mathbf{A}^{\mathrm{H}}\mathbf{D}\mathbf{z}_{k+1}-\frac{1}{e}(\boldsymbol{\mathrm{X}-}e\boldsymbol{\mathrm{I}})^{\mathrm{H}}\mathbf{w}_{k}$,
the above problem also becomes separable in $w_{l}$ 
\begin{equation}
\small\underset{|w_{l}|\leq1}{\mathrm{min}}\;\stackrel[l=1]{N}{\sum}|w_{l}|^{2}-2\mathrm{Re}\stackrel[l=1]{N}{\sum}(u_{l}^{*}w_{l}).
\end{equation}
Therefore, a generic problem can be written as 
\begin{equation}
\small\underset{|w|\leq1}{\mathrm{min}}\;|w|^{2}-2\mathrm{Re}\left(u^{*}w\right).\label{eq:62}
\end{equation}
The problem (\ref{eq:62}) has a closed form solution and the optimal
value of $w$ is given as
\begin{equation}
\small w_{k+1}=\begin{cases}
u\; & \mathrm{if}\;|u|\leq1\\
u/|u|\; & i\mathrm{f}\;|u|>1
\end{cases}\label{eq:63}
\end{equation}
The pseudo code for the MM algorithm for the regularized problem is
given in Algorithm Table \ref{alg:2}.

\begin{algorithm}[tbh]
\caption{\label{alg:2}Pseudo code of PDMM (regularized)}

Input: $\boldsymbol{\mathrm{A}}$, $\boldsymbol{\mathrm{T}}$, $\boldsymbol{\mathrm{y}}$,
$\boldsymbol{\mathrm{b}}$, $\eta_{1}$ and $\eta_{2}$
\begin{enumerate}
\item Initialize $\mathbf{x}_{0}$ and $\mathbf{z}_{0}$ and $\mathbf{w}_{0}$
\item Compute $\mathbf{A}^{\dagger}$, $\mathbf{P}$, $\mathbf{X}$, $\mathbf{(\mathrm{\mathbf{A}^{H}\mathrm{\mathbf{A}}})^{-1}\mathbf{T}^{\mathrm{H}}}$,
$\mathbf{\mathbf{A}(\mathbf{A}^{H}\mathrm{\mathbf{A}})^{-1}\mathbf{T}^{\mathrm{H}}}$
and $e$
\item Iterate: Given $\mathbf{x}_{t}$ , do the $(t+1)^{\mathrm{th}}$ step
\begin{enumerate}
\item Compute $\mathbf{d}$
\item Iterate: Given $\mathbf{w}_{k}$ and $\mathbf{z}_{k}$ , do the $(k+1)^{\mathrm{th}}$
step:
\begin{itemize}
\item Apply (\ref{eq:58}) to obtain $\mathbf{z}_{k+1}.$
\item Given $\mathbf{z}_{k+1}$, solve (\ref{eq:63}) for $\mathbf{w}_{k+1}.$
\item If $\left\Vert \mathbf{z}_{k+1}-\mathbf{z}_{k}\right\Vert _{2}/\left\Vert \mathbf{z}_{k}\right\Vert _{2}<\eta_{1},$
stop and return $\mathbf{z}_{k+1}$and $\mathbf{w}_{k+1}$
\end{itemize}
\item Apply (\ref{eq:50}) to obtain $\mathbf{x}_{t+1}$
\item If $\left\Vert \mathbf{x}_{t+1}-\mathbf{x}_{t}\right\Vert _{2}/\left\Vert \mathbf{x}_{t}\right\Vert _{2}<\eta_{2}$,
stop and return $\mathbf{x}_{t+1}.$
\end{enumerate}
\item $\mathbf{x}^{\mathrm{opt}}$ is the value of $\boldsymbol{\mathrm{x}}$
returned at convergence of the outer loop.
\end{enumerate}
\end{algorithm}

\subsection{Computational Complexity of PDMM}

For the unregularized problem, the main computational overheads of
PDMM are in the calculation of pseudo inverse $\mathbf{A}^{\dagger}$
(complexity $O(K^{2}N)$) and the projection matrix $\mathbf{P}$
(complexity of $O(N^{2}K)$), assuming $N>K$, both of which are calculated
outside the loops. Thus the per iteration computational complexity
of PDMM is dominated only by some matrix-vector multiplications (with
a worst case complexity $O(NK)$) and Hadamard products of vectors
(with a complexity $O(N)$) in the calculation of vectors $\mathbf{c}$,
$\mathbf{d}$, $\mathbf{z}$ and $\mathbf{x}$.

For the regularized problem, the pseudo inverse matrix $\mathbf{A}^{\dagger},$
the projection matrix $\mathbf{P}$ as well as the matrices $(\mathrm{\mathbf{A}^{H}\mathrm{\mathbf{A}}})^{-1}\mathbf{T}^{\mathrm{H}}$
(complexity of $O(K^{2}N)$), $\mathbf{A}(\mathrm{\mathbf{A}^{H}\mathrm{\mathbf{A}}})^{-1}\mathbf{T}^{\mathrm{H}}$
(complexity of $O(K^{2}N)$), $\mathbf{X}$ (complexity of $O(K^{2}N)$)
and maximum eigenvalue $e$ (complexity of $O(K^{3})$) are calculated
and stored outside the loops. For large dimensional problems, instead
of the eigenvalue $e$, the trace of the matrix $\mathbf{X}$ can
be used which would result in a looser upperbound and a slower converging
algorithm; however, would reduce the overall computational complexity.
The per-iteration computational complexity in the regularized PDMM
algorithm is only dominated by matrix-vector multiplication (with
a worst case complexity $O(NK)$) and Hadamard products of vectors
(with a complexity $O(N)$) in the calculation of vectors $\mathbf{c}$,
$\mathbf{d}$, $\mathbf{g}$, $\mathbf{z}$, $\mathbf{w}$ and $\mathbf{x}$.
The complexity in the regularized case is greater than the unregularized
case because of the calculation of more number of parameters.

In the case of a DFT matrix setting, none of the quantities are pre-calculated
and stored, but are invoked inside the loops along with other calculations
using Fast Fourier Transform (FFT) and Inverse Fast Fourier Transform
(IFFT) which further reduces the computational complexity of the algorithm.

\subsection{Convergence Analysis of PDMM}

In this sub-section, we will prove that the iterative steps of PDMM
always converges to a stationary point of the Poisson likelihood problem.
Since PDMM is a double loop MM algorithm, we will prove the convergence
of both the MM updates separately. Moreover, the convergence of the
MM update over primal variable $\mathbf{x}$ depends on the convergence
of the MM update over dual variable $\mathbf{z}$ (minimization problem
in (\ref{eq:34})). Therefore, we first prove the convergence of the
MM update over dual variable $\mathbf{z}$.

Let us rewrite the objective function of the dual problem in (\ref{eq:34})
as:
\begin{equation}
\begin{aligned}\small p(\mathbf{z})=\mathbf{z}^{\mathrm{H}}\boldsymbol{\mathrm{D}}^{\mathrm{H}}\mathbf{A}(\mathrm{\mathbf{A}^{H}\mathrm{\mathbf{A}}})^{-1}\mathbf{\mathbf{A}^{\mathrm{H}}}\mathbf{Dz}+\stackrel[i=1]{\mathrm{\mathit{N}}}{\sum}z_{i}b_{i}\\
-\stackrel[i=1]{\mathrm{\mathit{N}}}{\sum}y_{i}\,\mathrm{log}\,(z_{i})-\stackrel[i=1]{\mathrm{\mathit{N}}}{\sum}z_{i}\left|d_{i}\right|^{2}
\end{aligned}
\end{equation}
As explained in subsection \ref{subsec:3.1}, the sequence of points
$\left\{ \mathbf{z}_{k}\right\} $ generated by MM update results
in a monotonically decreasing objective function ($p(\mathbf{z})$).
It can be seen that $p(\mathbf{z})$ is convex and a continuous function,
and also bounded from below since for finite values of $y_{i}$, $b_{i}$
and $d_{i}$, the objective function $p(z_{i})\rightarrow\infty$
at the extremum points of the function i.e. $z_{i}\rightarrow\infty$
and/or $z_{i}\rightarrow0$. Therefore, the sequence generated by
MM algorithm $\left\{ p\left(\mathbf{z}_{k}\right)\right\} $ converges
to some finite value. From (\ref{eq:14}) we have 
\[
\small p\left(\mathbf{z}_{0}\right)\geq p\left(\mathbf{z}_{1}\right)\geq p\left(\mathbf{z}_{2}\right)\ldots
\]
We assume that there exists a convergent subsequence $\left\{ \mathbf{z}_{k_{j}}\right\} $
which converges to the limit point $\mathbf{z}_{\mathrm{limit}}$.
We next prove that $\mathbf{z}_{\mathrm{limit}}$ is a stationary
point. Since the dual problem is a constrained problem, for a point
$\mathbf{z}_{\mathrm{limit}}$ to be stationary, it needs to satisfy:
\[
\small\mathrm{Re}(\nabla p(\mathbf{z}_{\mathrm{limit}};\mathbf{d})^{\mathrm{H}}\mathbf{t})\geq0\;\forall\mathbf{t}\in T_{\mathbb{R}_{+}^{n}}(\mathbf{z}_{\mathrm{limit}})
\]
 where $T_{\mathbb{R}_{+}^{n}}(\mathbf{z}_{\mathrm{limit}})$ denotes
the tangent cone of $\mathbb{R}_{+}^{n}$(constraint set of $\mathbf{z}$)
at $\mathbf{z}_{\mathrm{limit}}$ and $\nabla p(\mathbf{z};\mathbf{d})$
is the directional derivative of the function $p(\mathbf{z})$ which
is defined as: 
\begin{equation}
\small\nabla p(\mathbf{z};\mathbf{d})=\underset{\alpha\rightarrow0}{\mathrm{lim}}\:\mathrm{inf}\:\frac{p\left(\mathbf{z}+\alpha\mathbf{d}\right)-p\left(\alpha\right)}{\alpha}
\end{equation}
Let $g_{p}\left(.\right)$ denote the surrogate function for the function
$p(\mathbf{z})$ (which is the objective of (\ref{eq:34})). Using
(\ref{eq:14}), we get the following inequality:
\begin{align}
\small\begin{aligned}g_{p}(\mathbf{z}_{k_{j+1}}|\mathbf{z}_{k_{j+1}})=p(\mathbf{z}_{k_{j+1}})\leq p(\mathbf{z}_{k_{j}+1})\end{aligned}
\\
\leq g_{p}(\mathbf{z}_{k_{j}+1}|\mathbf{z}_{k_{j}})\leq g_{p}(\mathbf{z}|\mathbf{z}_{k_{j}})
\end{align}
For a limit point, $j\rightarrow\infty$, and we get:
\begin{equation}
\small g_{p}\left(\mathbf{z}_{\mathrm{limit}}|\mathbf{z}_{\mathrm{limit}}\right)\leq g_{p}\left(\mathbf{z}|\mathbf{z}_{\mathrm{limit}}\right)
\end{equation}
which implies 
\[
\small\mathrm{Re}\left(\nabla g_{p}(\mathbf{z}_{\mathrm{limit}})^{\mathrm{H}}\mathbf{t}\right)\geq0\;\forall\mathbf{t}\in T_{\mathbb{R}_{+}^{n}}(\mathbf{z}_{\mathrm{limit}})
\]
Since the first order behavior of $p(\mathbf{z})$ is same as that
of $g_{p}(\mathbf{z})$, we get
\[
\small\mathrm{Re}\left(\nabla p(\mathbf{z}_{\mathrm{limit}})^{\mathrm{H}}\mathbf{t}\right)\geq0\;\forall\mathbf{t}\in T_{\mathbb{R}_{+}^{n}}(\mathbf{z}_{\mathrm{limit}})
\]
 thereby proving that $\mathbf{z}_{\mathrm{limit}}$ is a stationary
point of $p(\mathbf{z})$ . Since $p(\mathbf{z})$ is a strongly convex
function, $\mathbf{z}_{\mathrm{limit}}$ would also be a global minimizer
of $p(\mathbf{z})$.

The convergence of the MM algorithm over primal variable $\mathbf{x}$
can be proved in a similar way. Let us recollect the objective function
of the primal problem as:
\begin{equation}
\small f(\mathbf{x})=\stackrel[i=1]{\mathrm{\mathit{N}}}{\sum}\left[\left|\boldsymbol{\mathrm{a}}_{i}^{\mathrm{H}}\mathbf{x}\right|^{2}+b_{_{i}}-y_{i}\,\mathrm{log}\,\left(\left|\boldsymbol{\mathrm{a}}_{i}^{\mathrm{H}}\mathbf{x}\right|^{2}+b_{i}\right)\right]
\end{equation}
The sequence of points $\left\{ \mathbf{x}_{t}\right\} $ generated
by MM algorithm monotonically decreases the objective function $f(\mathbf{x})$
monotonically. Also, the objective function $f(\mathbf{x})$ is continuous
and bounded from below since for a finite value of $y_{i}$, the objective
function $f(\mathbf{x})\rightarrow\infty$ at the extremum points
of the function i.e. $x_{i}\rightarrow\infty$ and/or $x_{i}\rightarrow-\infty$,
for any $i$. Therefore, the sequence generated by MM algorithm $\left\{ f\left(\mathbf{x}_{t}\right)\right\} $
converges to some finite value at the limit point $\mathbf{x}_{\mathrm{limit}}$.
We next prove that $\mathbf{x}_{\mathrm{limit}}$ is a stationary
point. Since the primal problem is an unconstrained problem, for a
point $\mathbf{x}$ to be stationary, $\nabla f(\mathbf{x};\mathbf{d})=0$.
Similar to the argument used in the proof over dual variable $\mathbf{z}$,
we get
\[
\small g_{f}\left(\mathbf{x}_{\mathrm{limit}}|\mathbf{x}_{\mathrm{limit}}\right)\leq g_{f}\left(\mathbf{x}|\mathbf{x}_{\mathrm{limit}}\right)
\]
where $g_{f}\left(.\right)$ denotes the surrogate function as in
(\ref{eq:23}). The above result implies $\nabla g_{f}(\mathbf{x}_{\mathrm{limit}})=0$.
Since the first order behavior of $f(\mathbf{x})$ is same as that
of $g_{f}(\mathbf{x})$, we get $\nabla f(\mathbf{x}_{\mathrm{limit}})=0$,
thereby proving that $\mathbf{x}_{\mathrm{limit}}$ is a stationary
point of (\ref{eq:8}). Thus, by proving the convergence of the MM
updates over both primal and dual variables, we establish the proof
of convergence of the proposed algorithm.

\section{NUMERICAL SIMULATIONS}

This section discusses the simulation details of the proposed algorithm
under two different experimental settings. In the first setting, the
matrix $\mathbf{A}$ is taken to be a complex random matrix with its
elements having independent random real and imaginary parts following
Uniform distribution in the interval $\left(0,1\right)$. In the second
setting, the matrix $\mathbf{A}$ is modelled using the Discrete Fourier
Transform (DFT) matrix. In case of random matrix setting, a random
signal $\mathbf{x}_{\mathrm{true}}\in\mathbb{C^{\mathrm{K}}}$, normalized
as $\mathbf{x}_{\mathrm{true}}/\left\Vert \mathbf{x}_{\mathrm{true}}\right\Vert _{2}$,
is taken as the original signal. Whereas, for the DFT matrix setting,
an image of size $K\times K$ is the original signal. The background
signal $\mathrm{\boldsymbol{b}}$ in both the settings is a constant
vector with values equal to $0.1$ for the case $b_{i}>0$. All the
measurements are considered to be corrupted with noise ($\mathbf{n})$
following Poisson distribution. The measurement vector $\boldsymbol{\mathrm{y}}$
is modelled as $\mathrm{\mathbf{y}=\left|\mathbf{A}\mathrm{\mathbf{x}}_{o}\right|^{2}+\mathbf{b}+\mathbf{n}}$.
The performance of the proposed algorithm (PDMM) is compared with
the competing algorithms such as WF, ADMM, and MM proposed in \cite{key-37}.
Since the authors of \cite{key-37} have already compared their proposed
algorithm with the GS algorithm and WF (Gaussian) algorithm and have
established that the Poisson phase-retrieval algorithms perform better
than the algorithms for Gaussian model in case of Poisson data model,
we do not include the comparison of PDMM with the Gaussian phase-retrieval
algorithms in this paper.

The experiments are conducted using MATLAB (R2018a) on a personal
computer with 1.7 GHz Intel(R) Core(TM) i5-4210U CPU and 16.00 GB
RAM.

\subsection{Initialization and convergence threshold}

As suggested by the authors in \cite{key-4}, the leading eigenvector
of the matrix $\mathrm{\mathbf{A}^{H}\,diag\,(\mathbf{y}-\mathbf{b})\mathbf{A}}$
is taken as an initial estimate $\mathbf{\tilde{x}}_{0}.$ To tackle
signals of arbitrary scale, the leading eigenvector obtained is scaled
by a constant given by:
\begin{align*}
\small\hat{\alpha} & =\mathrm{arg}\:\underset{\alpha\in\mathbb{R}}{\mathrm{min}}\left\Vert \mathbf{y}-\mathbf{b}-\left|\alpha\mathbf{A}\tilde{\mathbf{x}}_{0}\right|{}^{2}\right\Vert _{2}^{2}\\
 & =\frac{\sqrt{\left((\mathbf{y}-\mathbf{b})^{\mathrm{T}}|\mathbf{A}\tilde{\mathbf{x}}_{0}|^{2}\right)}}{\left\Vert \mathbf{A}\tilde{\mathbf{x}}_{0}\right\Vert _{4}^{2}}.
\end{align*}
Therefore, the initial estimate $\mathbf{x}_{0}$ is $\hat{\alpha}\mathbf{\tilde{x}}_{0}$.
The vector $\mathbf{z}$ is initialized as $z_{i}=\frac{y_{i}}{\left|\boldsymbol{\mathrm{a}}_{i}^{\mathrm{H}}\mathbf{x}_{0}\right|^{2}+b_{i}}$
and vector $\mathbf{w}$ is initialized as a random unit vector once
outside the iteration loops.

The threshold $\eta_{1}$ for terminating the outer loop ($\frac{\left\Vert \mathbf{x}_{t+1}-\mathbf{x}_{t}\right\Vert _{2}|}{\left\Vert \mathbf{x}_{t}\right\Vert _{2}}<\eta_{1}$)
is fixed at $10^{-6}$ for every iteration. The threshold $\eta_{2}$
for terminating the the inner loop ($\frac{\left\Vert \mathbf{z}_{k+1}-\mathbf{z}_{k}\right\Vert _{2}}{\left\Vert \mathbf{z}_{k}\right\Vert _{2}}<\eta_{2}$)
can either be fixed and initialized outside the inner loop or can
be adaptively changed after each iteration by checking for a decrease
in the original objective function. (Please see the discussion in
Remark \ref{rem:4}).

\subsection{Ambiguities and Performance Evaluation}

Due to loss of global phase information, the phase-retrieval algorithms
can recover the original signal only within a constant phase shift.
Therefore in the case of random matrix setting, the following Normalized
Root Mean Square Error (NRMSE) is used to evaluate the performance
of proposed algorithm taking into consideration the global phase shift.
\begin{equation}
\small\mathrm{NRMSE}=\frac{\left\Vert \mathbf{x}^{\mathrm{opt}}-\mathbf{x}_{\mathrm{true}}\mathrm{e}^{\mathrm{i\phi}}\right\Vert _{2}}{\left\Vert \mathbf{x}_{\mathrm{true}}\right\Vert _{2}},\;\mathrm{e}^{\mathrm{i\phi}}=\mathrm{sign}(\mathbf{x}_{\mathrm{true}}^{'}\mathbf{x}^{\mathrm{opt}}),
\end{equation}
 where $\mathbf{x}^{\mathrm{opt}}$ denotes the recovered signal and
$\mathbf{x}_{\mathrm{true}}$ denotes the original signal.

In case of DFT Matrix setting, there are more number of ambiguities
like global constant phase shift, circular shift, conjugate inversion
and their combinations that conserve the Fourier magnitude and contribute
to trivial ambiguities. Also, two signals with the same autocorrelation
function have the same Fourier magnitude. This results in the recovery
of original signal only up to the same autocorrelation function without
any additional information. One method to deal with these ambiguities
is to use the mean squared error between the auto-correlation function
of the original and recovered signal for performance evaluation instead
of using NRMSE. Another method is to introduce redundancy in the measurement
vector $\mathbf{y}$ using masked DFT. Here, instead of the measurements
being made as:
\[
\small y_{n}=\left|\stackrel[k=0]{K-1}{\sum}x_{k}\mathrm{e}^{-i2\pi nk/N}\right|^{2}+b_{n},
\]
where $N=2K-1$, $M$ redundant masks $\mathbf{D}^{m}\:(m=1,...,M)$
are introduced and the measurement model becomes:
\[
\small y_{n}^{m}=\left|\stackrel[k=0]{K-1}{\sum}x_{k}D_{k}^{m}\mathrm{e}^{-i2\pi nk/N}\right|^{2}+b_{n}^{m}.
\]
Similar to \cite{key-37}, a total of $M=21$ masks (where the first
mask is a full sampling mask and the rest have a sampling rate of
0.5 with random sampling patterns) are used in the experiments to
define the measurement matrix $\mathbf{A}\in\mathbb{C}^{MN\times K}$.

\subsection{Experimental Settings}

This sub-section explains the details of different experiments performed
under both random matrix Setting and DFT matrix setting.

\subsubsection{Random Matrix Setting}

Under random matrix setting, three different experiments are performed.
In the first experiment, the length of $\mathbf{x}_{\mathrm{true}}$
is fixed at $K=100$ and the average running time and NRMSE of the
algorithms are plotted against the number of measurements $N$, where
$N$ is varied between $1000$ and $8000$ at an interval of $1000$.
In the second experiment, the number of measurements is fixed at $N=4000$
and the average running time and NRMSE of the algorithms are plotted
against the length of $\mathbf{x}_{\mathrm{true}}$ ($K$), where
$K$ is varied between $100$ and $500$ at an interval of $100.$
In the third experiment, with the number of measurements fixed at
$N=4000$ and the length of $\mathbf{x}_{\mathrm{true}}$ fixed at
$K=300$, the NRMSE is plotted against time. The plots of NRMSE against
$N$ and $K$ gives the analysis of the accuracy of the proposed algorithm,
whereas the plots of average running time against $N$ and $K$ gives
an analysis of the convergence of the proposed algorithm. The NRMSE
vs time plot gives an analysis of both accuracy and convergence. The
experiments are repeated for 50 Monte-Carlo simulations to calculate
the average computational time and NRMSE values. All the experiments
are performed for both when the background signal $b_{i}=0$ and $b_{i}>0.$

\subsubsection{DFT Matrix Setting}

In many phase retrieval problems such as those arising in optical
imaging, the measurements are the magnitudes of the Fourier transform.
The measurement matrix $\mathbf{A}$ in such cases is modelled as
a DFT matrix. Therefore, in the second experimental setting, we test
the proposed algorithm on an image of size $K\times K$, where the
measurement matrix $\mathbf{A}$ is as described in the previous sub-section
with $M=21$ masked DFT matrices. The matrix $\mathbf{A}$ is normalized
such that the average of $|\mathbf{a}_{i}^{\mathrm{H}}\mathbf{x}|^{2}$
is 1 for $i=1,...,N$. The Cameraman image of size $128\times128$
pixels is used as the original image. To demonstrate that PDMM can
be adapted to solve $\ell_{1}$ regularized problems, where the matrix
$\mathbf{T}$ may not be proximal friendly, the experiments under
DFT matrix setting are performed for TV regularized Poisson phase
retrieval problem. The reconstructed image along with the corresponding
NRMSE is compared with the original image. Furthermore, NRMSE is plotted
against time for the analysis of the convergence of the PDMM. The
results of PDMM is also compared with ADMM-TV and MM-TV algorithms
as proposed in \cite{key-37} which use alternating minimization and
Conjugate Gradient method in their iterations to solve for the optimal
value of $\mathbf{x}$. The WF algorithm proposed in \cite{key-37}
is not considered for comparison in the case of DFT matrix setting
because gradient based methods like WF are not suitable for non-smooth
$\ell_{1}$ regularizers. Similar to \cite{key-37}, the value of
regularization parameter $\left(\lambda\right)$ is chosen to be 8.

\subsection{SIMULATION RESULTS}

This section gives the simulation results of the experiments for both
random matrix setting and DFT matrix setting and compares the result
with the competing algorithms such as the WF, MM and ADMM algorithms
for Poisson distribution as proposed in \cite{key-37}.

\subsubsection{Random Matrix Setting}
\begin{enumerate}
\item Average time vs Number of measurements ($N$)

Fig. \ref{fig:2} shows the comparison of average running time against
the number of measurements ($N$) for all algorithms. Sub-Figures
(a) and (b) show results for cases when background signal $b_{i}=0.1$
and $b_{i}=0$ respectively. It is observed that PDMM is faster than
the competing algorithms.
\begin{center}
\begin{figure}[tbh]
\centering{}\subfloat[]{\begin{centering}
\includegraphics[width=6.8cm,height=3.6cm]{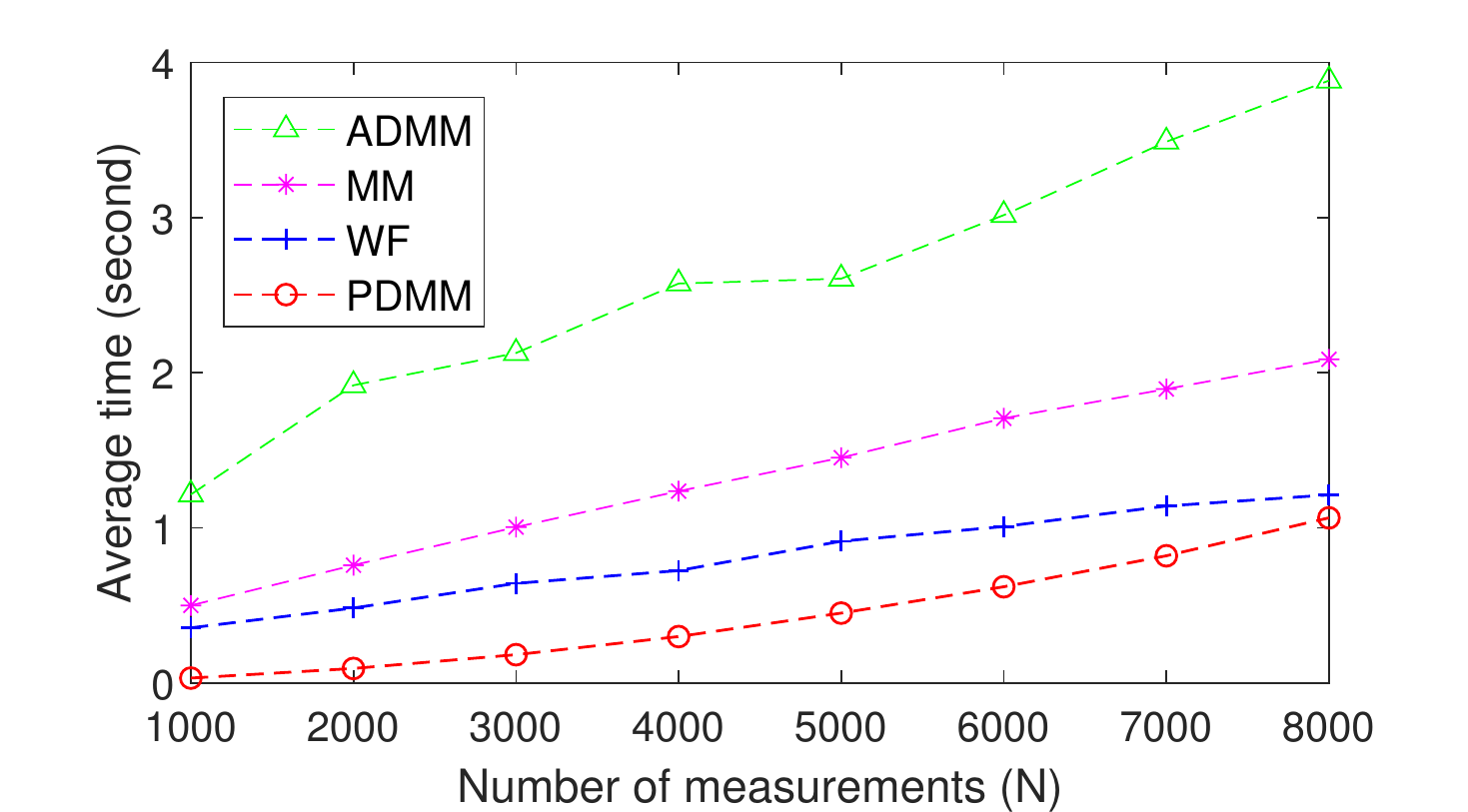}
\par\end{centering}
}\vfill{}
\subfloat[]{\begin{centering}
\includegraphics[width=6.8cm,height=3.6cm]{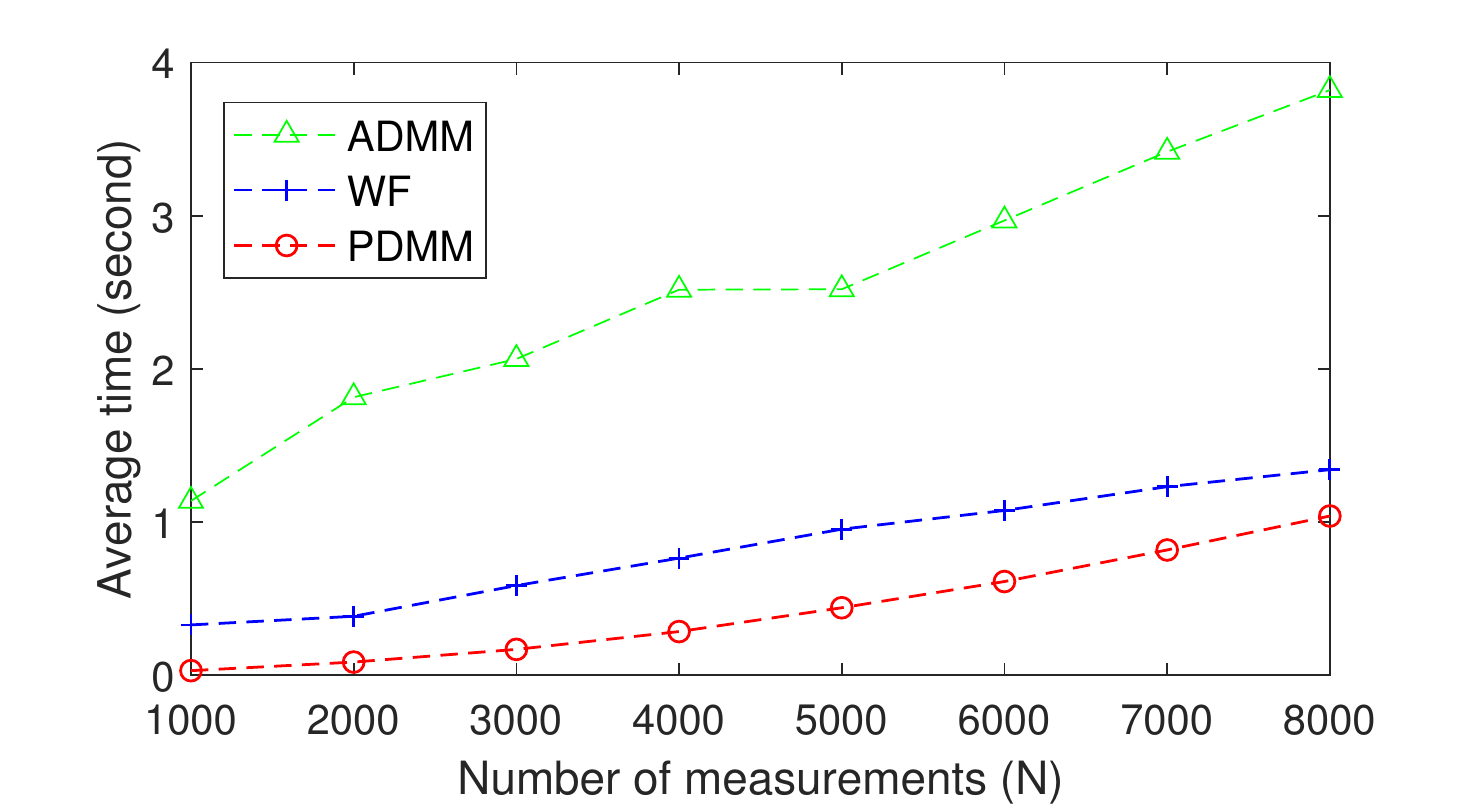}
\par\end{centering}
}\caption{\label{fig:2}Average computation time (in seconds) vs Number of measurements
$N$, when $\eta=10^{-6}$, $\mathbf{x}\in\mathbb{C}^{100}$ and $\mathbf{A}$
is a random matrix. Sub-Figures (a) and (b) corresponds to cases when
$b_{i}=0.1$ and $b_{i}=0$ respectively.}
\end{figure}
\par\end{center}
\item NRMSE vs Number of measurements ($N$)

Fig. \ref{fig:3} gives the comparison of the NRMSE against the number
of measurements ($N$) for all algorithms. The plot for PDMM overlaps
almost completely with the WF, ADMM and MM algorithms. Thus, in terms
of the accuracy of the recovered signal, the performance of PDMM is
at par with the previously proposed algorithms.
\begin{center}
\begin{figure}[tbh]
\centering{}\subfloat[]{\begin{centering}
\includegraphics[width=6.8cm,height=3.6cm]{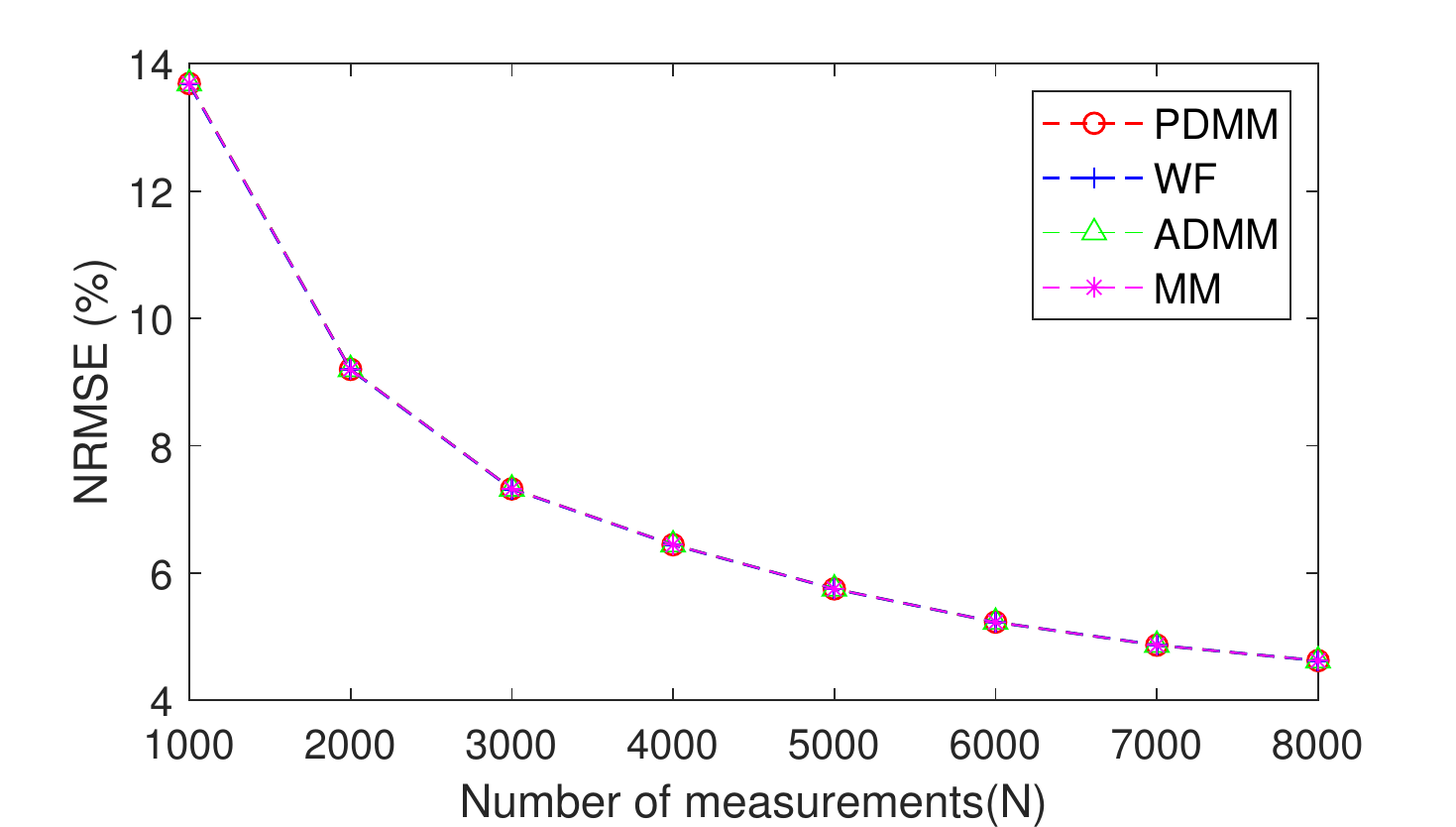}
\par\end{centering}
}\vfill{}
\subfloat[]{\begin{centering}
\includegraphics[width=6.8cm,height=3.6cm]{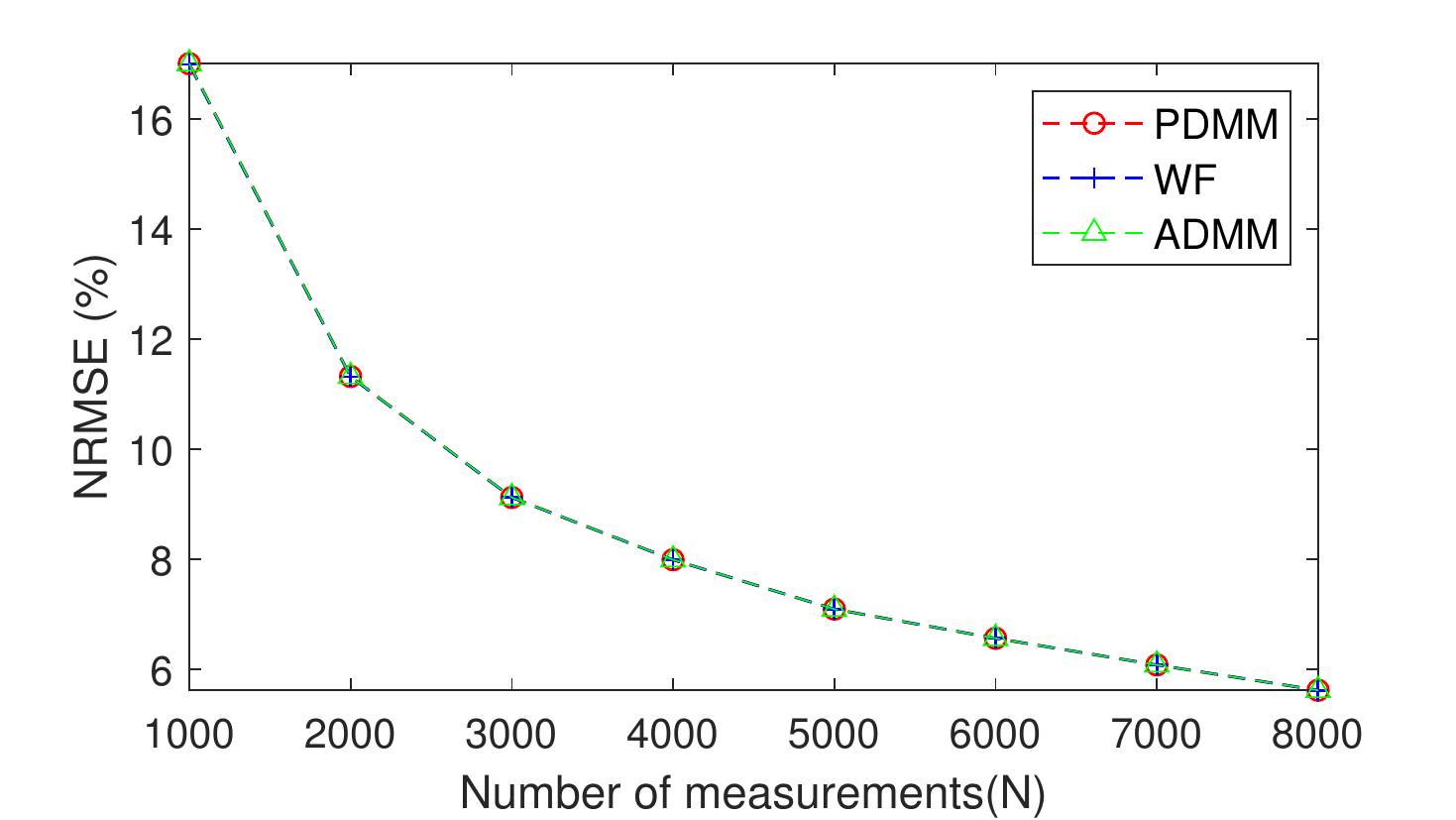}
\par\end{centering}
}\caption{\label{fig:3}NRMSE vs Number of measurements ($N$) for $\eta=10^{-6}$
, $\mathbf{x}\in\mathbb{C}^{100}$ and $\mathbf{A}$ is a random matrix.
Sub-Figure (a) and (b) corresponds to the cases when $b_{i}=0.1$
and $b_{i}=0$ respectively.}
\end{figure}
\par\end{center}
\item Average time vs Length of the original signal ($K$)

Fig. \ref{fig:4} compares the plots of average running time against
the length of the original signal ($K$) for all algorithms. For a
fixed number of measurements, PDMM is the fastest for different lengths
of $\mathbf{x}_{\mathrm{true}}$. The computational complexity of
the MM algorithm increases with the increase in the length of $\mathbf{x}_{\mathrm{true}}$
perhaps due to the calculation of inverse of a $K\times K$ matrix
performed at every iteration of the algorithm. This can be reduced
by using CG for updating $\mathbf{x}.$ The WF algorithm converges
slowly for larger values of $K.$
\begin{center}
\begin{figure}[tbh]
\centering{}\subfloat[]{\begin{centering}
\includegraphics[width=6.8cm,height=3.6cm]{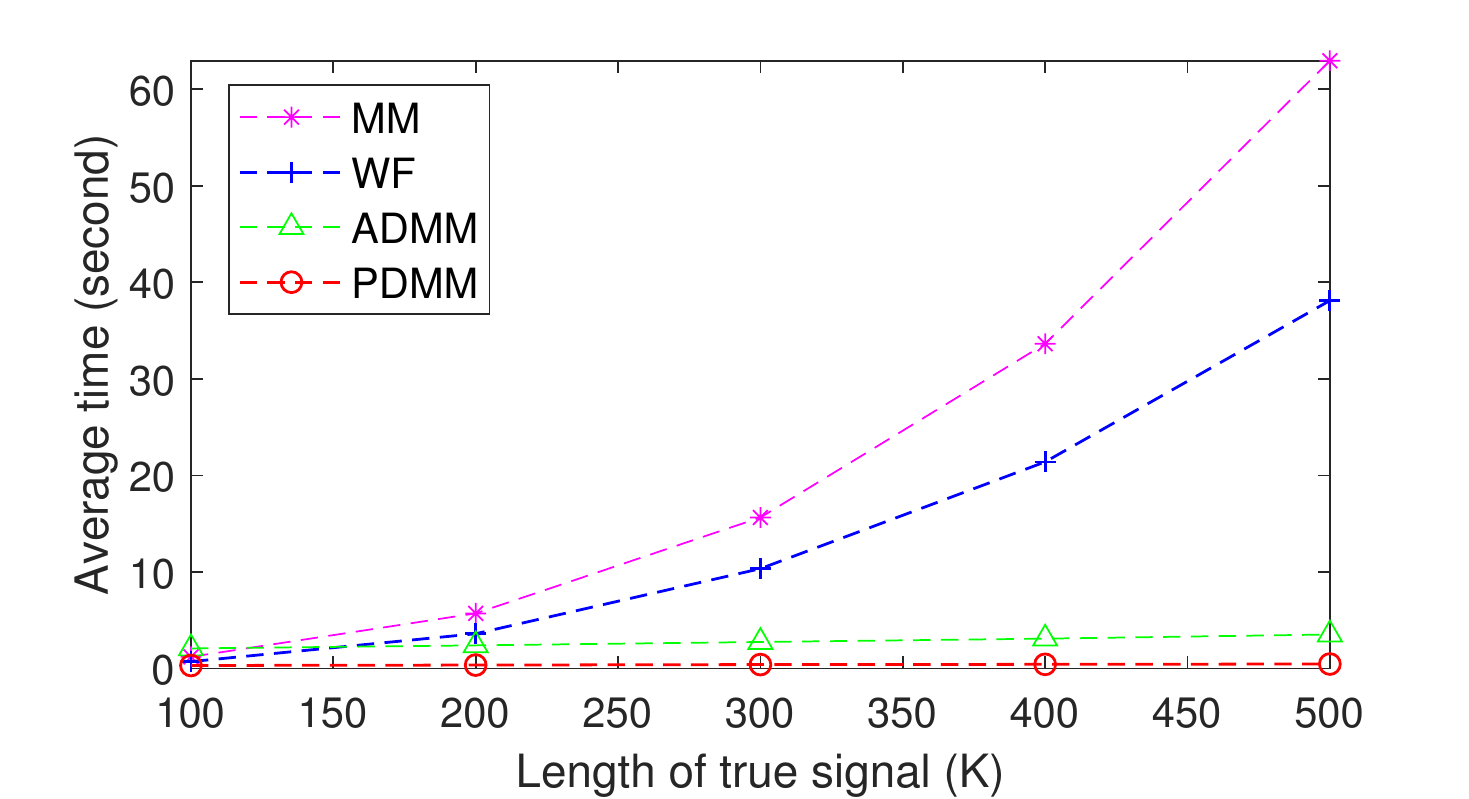}
\par\end{centering}
}\vfill{}
\subfloat[]{\begin{centering}
\includegraphics[width=6.8cm,height=3.6cm]{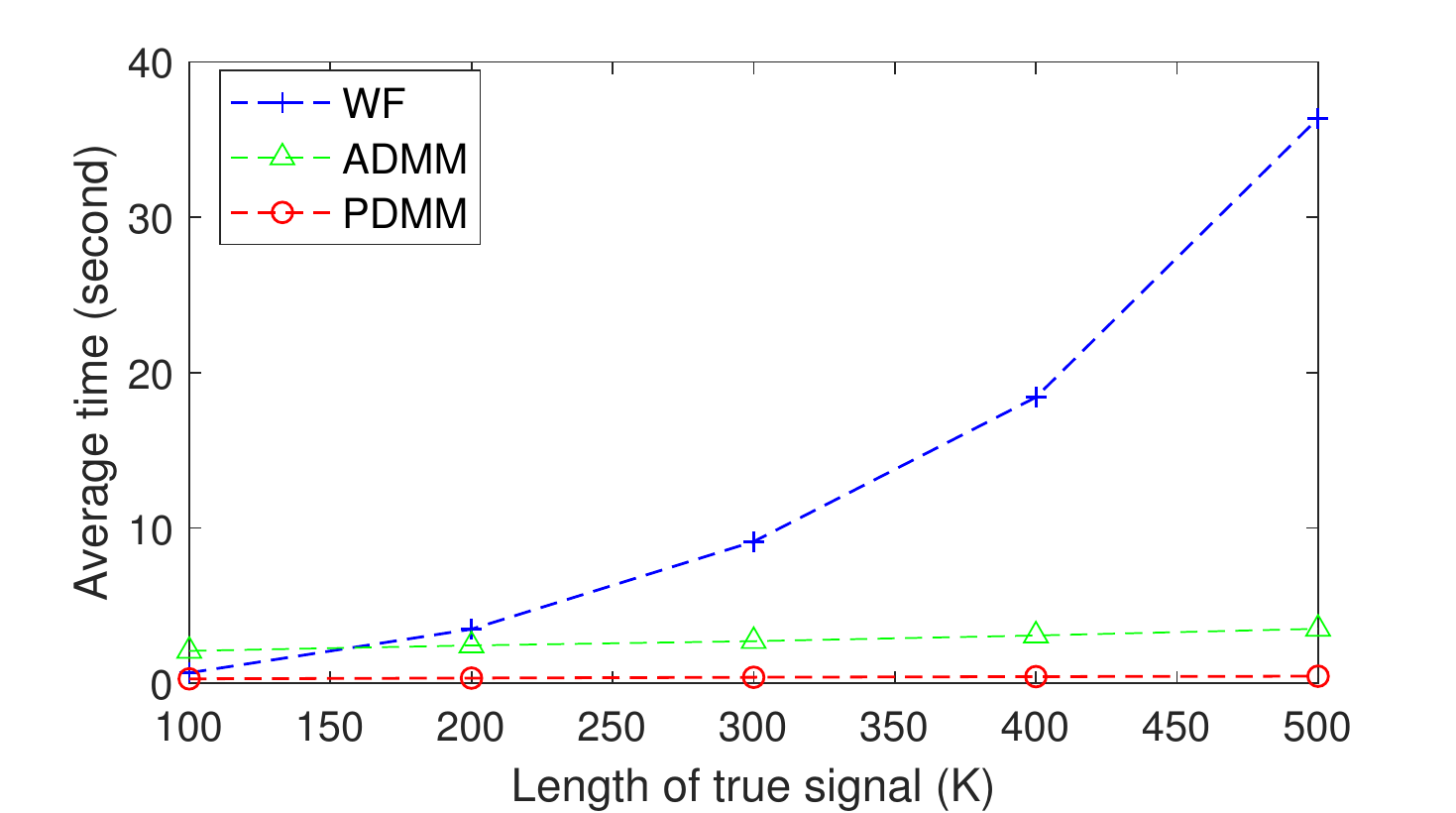}
\par\end{centering}
}\caption{\label{fig:4}Average computation time (in seconds) vs Length of original
signal $K$, when $\eta=10^{-6}$, $N=4000$ and $\mathbf{A}$ is
a random matrix. Sub-Figures (a) and (b) corresponds to cases when
$b_{i}=0.1$ and $b_{i}=0$ respectively.}
\end{figure}
\par\end{center}
\item NRMSE vs Length of original signal ($K$)

Fig. \ref{fig:5} gives the comparison of the plots of NRMSE against
the length of $\mathbf{x}_{\mathrm{true}}$ ($K$) for each algorithm.
As expected, the NRMSE increases with $K$ because of increase in
the number of parameters estimated. The plot for PDMM overlaps almost
completely with WF, ADMM and MM algorithm indicating its performance
to be at par with the previously proposed algorithms in terms of successful
recovery of the original signal.
\begin{center}
\begin{figure}[tbh]
\centering{}\subfloat[]{\begin{centering}
\includegraphics[width=6.8cm,height=3.6cm]{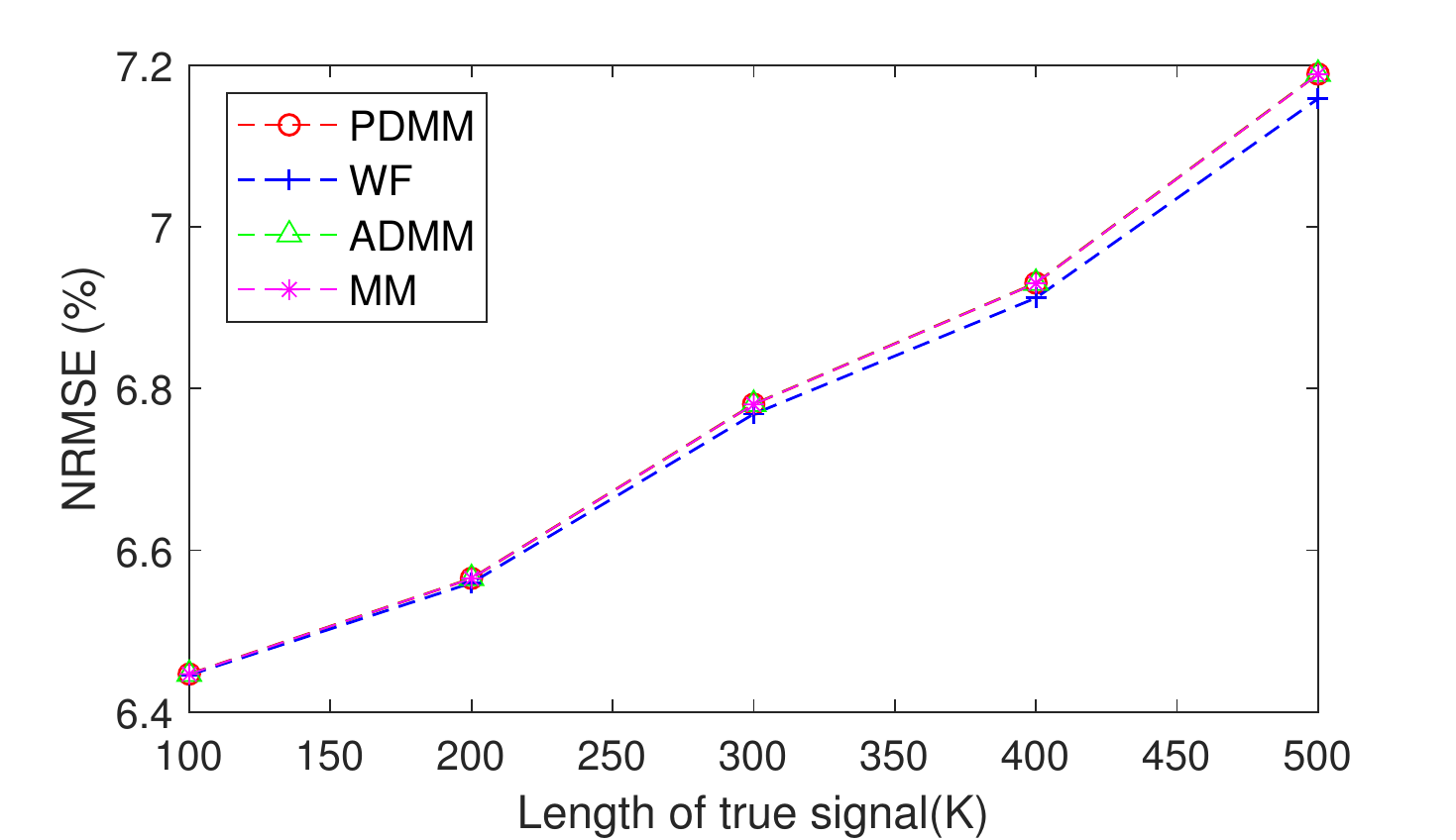}
\par\end{centering}
}\vfill{}
\subfloat[]{\begin{centering}
\includegraphics[width=6.8cm,height=3.6cm]{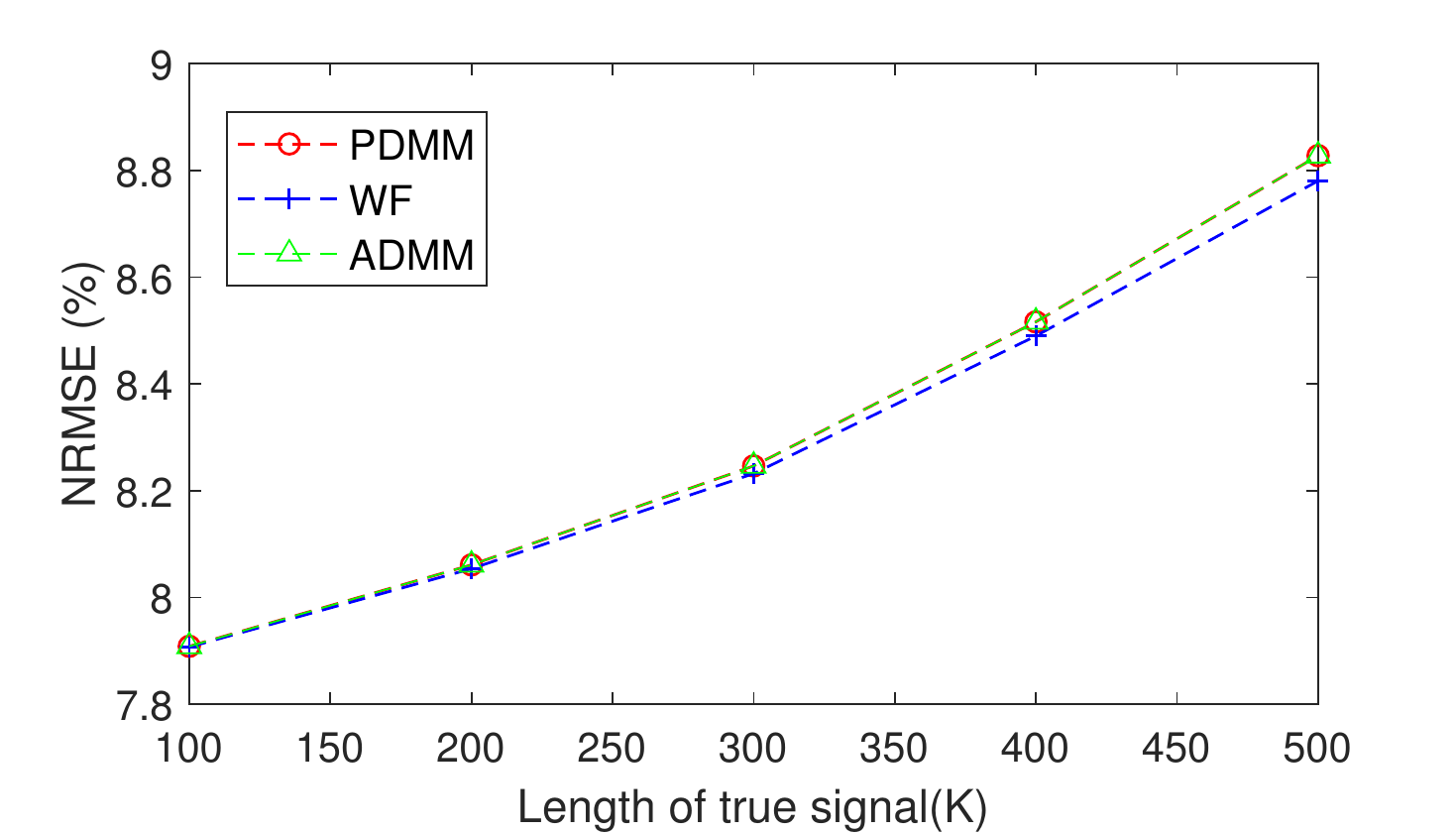}
\par\end{centering}
\centering{}}\caption{\label{fig:5}NRMSE vs Length of original signal ($K$) for $\eta=10^{-6}$
, $N=4000$ and $\mathbf{A}$ being a random matrix. Sub-Figure (a)
and (b) corresponds to the cases when $b_{i}=0.1$ and $b_{i}=0$
respectively.}
\end{figure}
\par\end{center}
\item NRMSE vs Time

Fig. \ref{fig:6} gives the NRMSE vs time plots for all the algorithms.
The number of measurements is fixed at $N=4000$ and the length of
$\mathbf{x}_{\mathrm{true}}$ is $300$. It is observed that PDMM
converges faster than the other three algorithms.
\end{enumerate}

\subsubsection{DFT Matrix Setting}

The experiments under DFT matrix setting were performed for the TV
regularized Poisson likelihood problem. Fig. \ref{fig:9} compares
the result of PDMM and other algorithms for Cameraman image of size
$128\times128$ pixels. The figure shows the original image along
with the recovered images using PDMM, ADMM and MM algorithm. The corresponding
NRMSE is mentioned below each image. Fig. \ref{fig:10} gives NRMSE
vs time plots for all the algorithms for the said image. The speed
of the proposed algorithm is comparable to the MM algorithm and is
much faster than the ADMM algorithm proposed in \cite{key-37}. In
terms of recovery of the original image, its performance is at par
with the state-of-the-art algorithms.

\section{CONCLUSION}

This paper introduces a novel method for Poisson phase-retrieval based
on the MM framework. In this method, using Fenchel representation
of the log term, an auxiliary dual variable is introduced and the
problem is converted into a saddle-point minimax problem. Surrogate
functions over both primal and dual variables are proposed, resulting
in a double loop MM algorithm. The resulting primal-dual majorization-minimization
(PDMM) algorithm is compared against the existing maximum likelihood
(ML) estimation algorithms for solving the Poisson phase-retrieval
problem. It is observed that the proposed algorithm (PDMM) is in general
faster than the algorithms proposed in \cite{key-37}. The performance
of PDMM in terms of the accuracy of recovered signal/image is at par
with previously proposed algorithms. Unlike the other algorithms,
the proposed algorithm can be easily adapted to regularized problem,
where the regularizer may not be smooth and proximal friendly. Furthermore,
the previously proposed MM algorithm, where a quadratic majorizer
is used, works only in cases where background signal $b_{i}>0$, whereas
PDMM works even when $b_{i}=0$.
\begin{center}
\begin{figure}[tbh]
\centering{}\subfloat[]{\begin{centering}
\includegraphics[width=6.8cm,height=3.6cm]{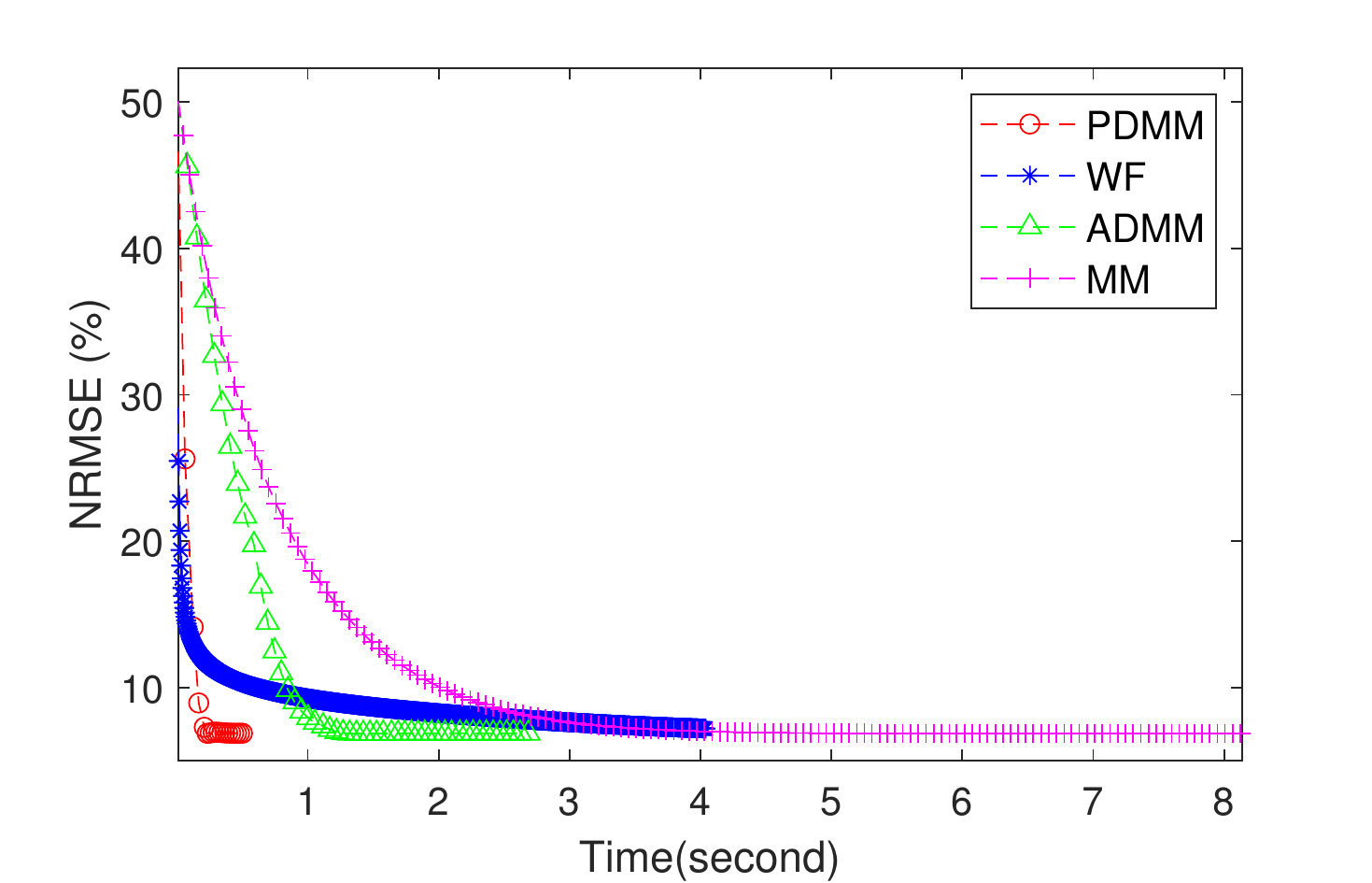}
\par\end{centering}
}\vfill{}
\subfloat[]{\begin{centering}
\includegraphics[width=6.8cm,height=3.6cm]{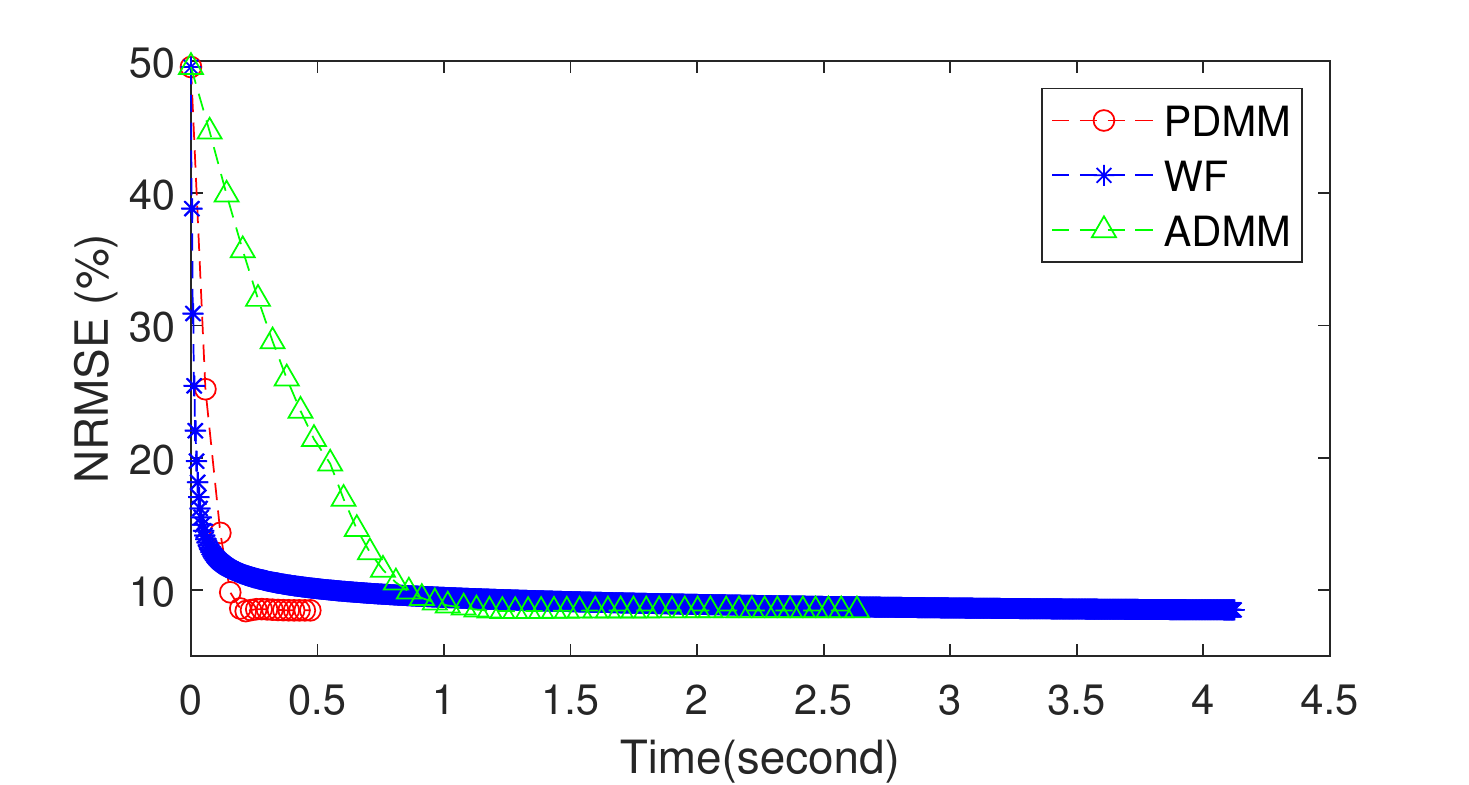}
\par\end{centering}
}\caption{\label{fig:6}NRMSE vs time when $N=4000$ , $\mathbf{x}\in\mathbb{C}^{300}$,
$\eta=10^{-6}$ and $\mathbf{A}$ is a random matrix. Sub-Figure (a)
and (b) corresponds to the cases when $b_{i}=0.1$ and $b_{i}=0$
respectively.}
\end{figure}
\begin{figure}[tbh]
\centering{}\subfloat[{\scriptsize{}True Image}]{\begin{centering}
\includegraphics[scale=0.4]{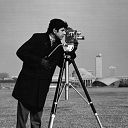}
\par\end{centering}
}\subfloat[{\scriptsize{}PDMM (6.3\%)}]{\begin{centering}
\includegraphics[scale=0.4]{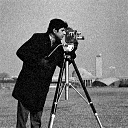}
\par\end{centering}
}\subfloat[{\scriptsize{}ADMM (6.9\%)}]{\begin{centering}
\includegraphics[scale=0.4]{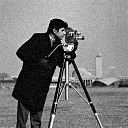}
\par\end{centering}
\centering{}}\subfloat[{\scriptsize{}MM (7.5\%)}]{\begin{centering}
\includegraphics[scale=0.4]{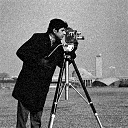}
\par\end{centering}
}\caption{\label{fig:9}Reconstructed image and corresponding NRMSE compared
to the true image (Cameraman of size $128\times128$), for a measurement
system with $M=21$ masked DFT matrices and TV regularized phase-retrieval
problem.}
\end{figure}
\begin{figure}[tbh]
\begin{centering}
\includegraphics[width=6.8cm,height=3.6cm]{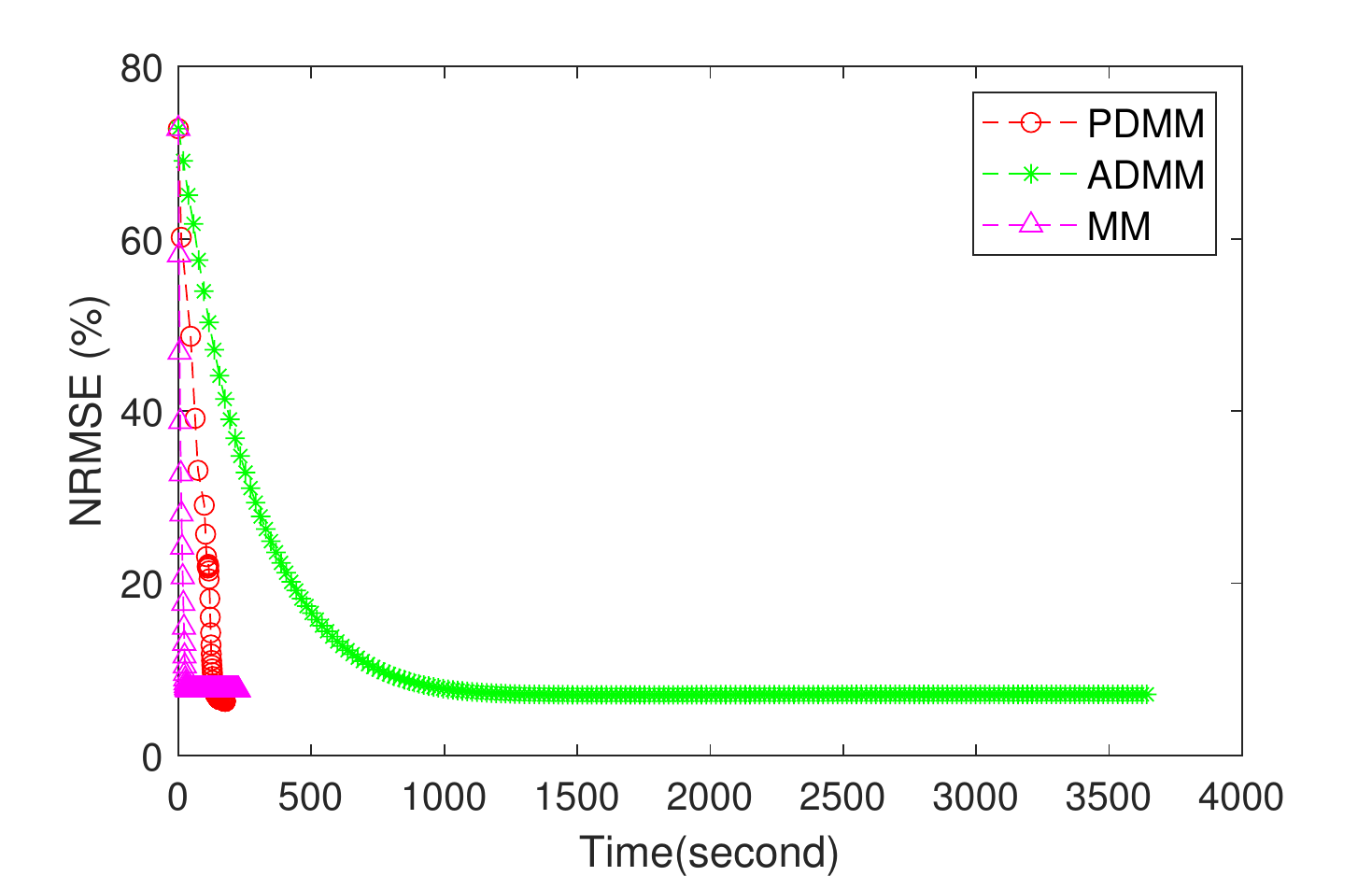}
\par\end{centering}
\caption{\label{fig:10}NRMSE vs time plot for a $128\times128$ Cameraman
image for a measurement system with $M=21$ masked DFT matrices and
TV regularized phase-retrieval problem.}
\end{figure}
\par\end{center}

\section*{ACKNOWLEDGEMENT}

We would like to thank Professor Jeffrey A. Fessler for his feedback
on the draft.

\end{document}